%% file: instantonSCFTs.tex
\documentclass[11pt,letterpaper]{article}
\RequirePackage{pdf14}
\usepackage[utf8]{inputenc}
\usepackage{graphicx,tikz}
\usepackage{amsfonts,amsmath,amssymb}
\usepackage{array,setspace,mathrsfs,yfonts,dsfont,bbm,colonequals,amscd,mathtools}
\usepackage{relsize,suffix,cancel,bbm,capt-of,upgreek,verbatim,xcolor}
\usepackage{dsfont}
\usepackage{jheppub}

\numberwithin{equation}{section}

\usepackage{amsmath}

\numberwithin{equation}{section}

\def\bea{\begin{eqnarray}}
\def\eea{\end{eqnarray}}

\def\ksu2{k_{2d}^{\suf(2)}}

\DeclarePairedDelimiterX\MeijerM[3]{\lparen}{\rparen}%
{\begin{smallmatrix}#1 \\ #2\end{smallmatrix}\delimsize\vert\,#3}

\newcommand\MeijerG[8][]{%
  G^{\,#2,#3}_{#4,#5}\MeijerM[#1]{#6}{#7}{#8}}

\WithSuffix\newcommand\MeijerG*[7]{%
  G^{\,#1,#2}_{#3,#4}\MeijerM*{#5}{#6}{#7}}

\def\bZ{\mathbb{Z}}

\def\cD{\mathcal{D}}

\def\fg{\mathfrak{g}}

\def\hv{{h^\vee}}

\def \beg#1{\begin{#1}} 

\def \bea{\beg{eqnarray}}
\def \eea{\end{eqnarray}}
\def \ee{\end{equation}}

\def \goodchi{\protect\raisebox{1pt}{$\chi$}}

\def \af{\mf{a}}
\def \df{\mf{d}}
\def \ef{\mf{e}}
\def \gf{\mf{g}}

\def \slf{\mf{sl}}
\def \suf{\mf{su}}

\def \ef{\mf{e}}

\def \restr#1#2{{\left.\kern-\nulldelimiterspace#1\vphantom{\big|}\right|_{#2}}}

\def \mf{\mathfrak}

\def \nn{\nonumber}

\def \eg{{\it e.g.}}

\def \Cb{\mathbb{C}}

\def \Hb{\mathbb{H}}

\def \Zb{\mathbb{Z}}

\def \CC{\mathcal{C}}

\def \JJ{\mathcal{J}}

\def \MM{\mathcal{M}}
\def \NN{\mathcal{N}}

\def \SS{\mathcal{S}}
\def \TT{\mathcal{T}}

\def \WW{\mathcal{W}}

\begin{document}

\title{VOAs and rank-two instanton SCFTs}

\author[\hspace{-.225em}1]{Christopher Beem,}
\author[\hspace{-.225em}1]{Carlo Meneghelli,}
\author[\hspace{-.225em}1]{Wolfger Peelaers,}
\author[\hspace{.08em}2,3]{Leonardo Rastelli}

\affiliation[1]{Mathematical Institute, University of Oxford, Woodstock Road, Oxford, OX2 6GG, United Kingdom}
\affiliation[2]{C. N. Yang Institute for Theoretical Physics, Stony Brook University, Stony Brook, NY 11794, USA}
\affiliation[3]{CERN, Theoretical Physics Department, 1211 Geneva 23, Switzerland}

\preprint{YITP-SB-19-20}

\abstract{We analyze the $\NN=2$ superconformal field theories that arise when a pair of D3-branes probe an F-theory singularity from the perspective of the associated vertex operator algebra. We identify these vertex operator algebras for all cases; we find that they have a completely uniform description, parameterized by the dual Coxeter number of the corresponding global symmetry group. We further present free field realizations for these algebras in the style of recent work by three of the authors. These realizations transparently reflect the algebraic structure of the Higgs branches of these theories. We find fourth-order linear modular differential equations for the vacuum characters/Schur indices of these theories, which are again uniform across the full family of theories and parameterized by the dual Coxeter number. We comment briefly on expectations for the still higher-rank cases.}

\maketitle
\setcounter{page}{1}

\section{Introduction and summary}
\label{sec:intro}

Four-dimensional $\NN=2$ superconformal field theories (SCFTs) showcase a remarkably rich diversity. Some admit Lagrangian descriptions, but many more are (conformal gaugings of) isolated, strongly coupled theories. A typical theory of class $\mathcal{S}$ is of the latter type \cite{Gaiotto:2009we,Gaiotto:2009hg}, as are all models of Argyres-Douglas kind.\footnote{While these theories do not admit a manifestly $\NN=2$ supersymmetric Lagrangian description, a variety of them have been argued to be the endpoints of Lagrangian $\NN=1$ renormalization group flows \cite{Maruyoshi:2016aim,Agarwal:2016pjo,Agarwal:2017roi}, or to lie on the $\NN=1$ conformal manifolds of Lagrangian theories \cite{Razamat:2019vfd}.} A useful characteristic by which one may organize this menagerie of theories is their \emph{rank}, \emph{i.e.}, the complex dimension of their Coulomb branch of vacua. A series of incrementally refined papers culminated in a conjectured classification and characterization of all rank-one theories \cite{Argyres:2015ffa,Argyres:2015gha,Argyres:2016xua,Argyres:2016xmc} (see also \cite{Caorsi:2019vex}). For higher ranks, a similar feat has not yet been achieved, though for partial progress see \cite{Argyres:2018zay,Caorsi:2018zsq}.

An interesting family of higher-rank theories are the rank-$n$ F-theory SCFTs, \emph{i.e.}, the low-energy worldvolume theories of stacks of $n$ D3-branes probing F-theory singularities \cite{Sen:1996vd,Banks:1996nj,Dasgupta:1996ij,Minahan:1996fg,Minahan:1996cj}. The possible choices of singularity follows the Kodaira classification, with the resulting interacting theories being labeled $H_0,H_1,H_2,D_4,E_6,E_7$, or $E_8$. Their flavor symmetries include as simple factors the corresponding simple Lie groups (with $H_i\rightarrow A_i$, and $H_0$ corresponding to the trivial Lie group), and, for $n>1$, also a factor of $SU(2)$. What's more, these theories have the beautiful property that their Higgs branches are the moduli spaces of $n$ centered $\gf$-instantons in $\mathbb{R}^4$, which is why they are sometimes referred to as rank-$n$ instanton SCFTs.\footnote{Here $\mathrm{Lie}(G)=\gf$, with $G$ the flavor symmetry factor just mentioned.} In addition, despite their uniform description in F-theory, for each $n$ this family of theories contains representatives of all three above-mentioned categories: one is Lagrangian ($D_4$), three admit class $\mathcal{S}$ descriptions ($E_6,E_7,E_8$), and the remaining three are of Argyres-Douglas type. The rank-two series of instanton SCFTs is the subject of interest in this paper.

A substantially more intricate invariant of four-dimensional $\NN=2$ SCFTs than their Coulomb branch is their associated vertex operator algebra (VOA), which arises by performing a cohomological truncation of the operator product algebra of local operators \cite{Beem:2013sza}.\footnote{We use the terms vertex operator algebra and chiral algebra interchangeably.} The VOA repackages an infinite amount of protected conformal data and as such provides an indispensable jumping-off point for a full analysis of the SCFT. The chiral algebras $\mathcal V_{\gf}^{(1)}$ associated with the rank-one $\gf$-instanton SCFTs have been identified in \cite{Beem:2013sza,Beem:2014rza,Buican:2015ina,Beem:2017ooy}. They admit a uniform description as $\hat\gf$ affine current algebras at level $k_{2d} = \frac{-\hv-6}{6}$, where $\hv$ denotes the dual Coxeter number of the Lie algebra $\gf$.\footnote{The $H_0$ F-theory SCFT is a special case. It has no flavor symmetry and its chiral algebra is the $(2,5)$ Virasoro VOA. Nevertheless, for many purposes it fits within the rank-one series upon formally setting $\hv=\frac{6}{5}$.} From the VOA viewpoint, there is no obstruction---and in fact it appears quite natural---to include two additional Lie algebras, $\gf_2$ and $\mf{f}_4$, to the previously listed seven, thus completing the so-called Deligne-Cvitanovi\'c series of exceptional Lie algebras \cite{Deligne,Cvitanovic:2008zz}. Their inclusion is suggested by the observation that the resulting series of nine current algebras are uniquely singled out as those whose levels and Virasoro central charges simultaneously saturate three independent (four-dimensional) unitarity bounds \cite{Beem:2013sza,Beem:2017ooy}. While their higher-rank cousins are not known to be singled out in such fashion, we find that the higher-rank VOAs still behave in a remarkably uniform fashion.

To understand the VOAs $\mathcal V_{\gf}^{(2)}$ associated with the rank-two $\gf$-instanton SCFTs, we pursue two different strategies. The first one is to set up and solve the bootstrap problem for these chiral algebras. The \textit{sine qua non} of this approach is a proposal for the list of strong generators of $\mathcal V_{\gf}^{(2)}$. Our Ansatz will be the minimal one compatible with general four-dimensional principles, consisting of only generators descending from Higgs branch chiral ring generators; these were shown to necessarily give rise to strong generators in \cite{Beem:2013sza}. Concretely, we set out to construct vertex operator algebras that are strongly generated by affine $\suf(2)$ and $\gf$ currents and a conformal weight $h=\tfrac{3}{2}$ generator transforming in the representation $(\tfrac{1}{2},\mathbf{Adj})$ of $\suf(2)\times \gf$.\footnote{The cases $H_0$ and $H_2$ behave slightly differently. The rank-two $H_0$ chiral algebra is generated by an affine $\suf(2)$ current algebra and an additional generator transforming as an $\suf(2)$ doublet and of conformal weight $\tfrac{5}{2}$. For $H_2$, on the other hand, we must include an independent Virasoro stress tensor, as the Sugawara construction fails to provide one due to the criticality of both affine current algebras.} We find that the associativity constraints can be solved uniquely and that the resulting OPE coefficients have a uniform expression in terms of $\hv$. As an aside, we note that the ``exotic'' SCFT dubbed $\mathcal{T}_X$ in \cite{Buican:2017fiq} can be recognized here as being exactly the rank-two $\mathfrak{a}_2$ instanton SCFT.

In principle there is no obstruction to pursuing this approach to construct the VOAs $\mathcal V_{\gf}^{(n)}$ for $n\geqslant3$, and we expect that they will all admit similarly uniform descriptions. However, the list of strong generators grows with $n$ because, on the one hand, generators descending from the Higgs branch chiral ring proliferate and, on the other hand, additional generators not related to Higgs branch chiral ring operators will make an appearance, rendering the bootstrap problem more involved. At the end of this paper we make a conjecture for the complete list of strong generators for $\mathcal V_{\gf}^{(n)}$ on the basis of a detailed analysis of the Schur limit of the superconformal index; we have not yet attempted to construct the corresponding VOAs.

Our second, complementary strategy is to realize these same vertex operator algebras using free fields, as proposed in \cite{Beem:2019tfp}. The low-energy degrees of freedom in a generic Higgs branch vacuum of a rank-two $\gf$-instanton SCFT consist of $2(2\hv-1)$ free half-hypermultiplets. Consequently, according to \cite{Beem:2019tfp}, we should anticipate the existence of (and could attempt to construct) a free field realization in terms of as many chiral bosons. However, we find a more economical approach by considering an intermediate, non-generic (but more symmetric) locus on the Higgs branch. Specifically, we consider the locus of the Higgs branch that preserves the full $G$-symmetry. A dense, open subset of this locus is isomorphic to $T^{*}(\Cb^\ast)$, where the residual interacting degrees of freedom at any point on this locus comprise two copies of the rank-one $\gf$-instanton SCFT. From this analysis we are led to a uniform free field construction in terms of two copies of the rank-one $\gf$-instanton VOA $\mathcal V_{\gf}^{(1)}$ accompanied by two chiral bosons. The success of this construction rests upon an exceptionally fine-tuned conspiracy of the various ingredients.\footnote{The simplest instance is the numerological fact that the affine level of the $\hat\gf$ current subalgebra of $\mathcal V_{\gf}^{(2)}$ equals twice the level of the $\hat\gf$ subalgebra of $\mathcal V_{\gf}^{(1)}$. This equality guarantees that the $\hat \gf$ current subalgebra of $\mathcal V_{\gf}^{(2)}$, realized as the diagonal sum of the $\hat\gf$ current algebras of the two copies of $\mathcal V_{\gf}^{(1)}$, has the correct level.} For example, non-trivial null relations for the two copies of $\mathcal V_{\gf}^{(1)}$ are required in order for the subspace of the free field state space that is strongly generated by the free field realized generators of $\mathcal V_{\gf}^{(2)}$ to be closed under the OPE. In a similar vein, these free field realizations are apparently simple (as modules over themselves), precisely because the two copies of $\mathcal V_{\gf}^{(1)}$ are already taken to be their simple quotients. Note that one could opt to realize each of the two copies of the vertex operator algebras $\mathcal V_{\gf}^{(1)}$ in terms of $2(\hv-1)$ chiral bosons using the construction of \cite{Beem:2019tfp}, and in doing so arrive at a construction of $\mathcal V_{\gf}^{(2)}$ in terms of the expected $2(2\hv-1)$ chiral bosons. The abstract algebras encoded in the free field realizations turn out to be identical to the results of the more direct, but technically more involved, chiral algebra bootstrap approach of the previous paragraph---all roads lead to Rome.

An important entry of the VOA/SCFT dictionary states that the vacuum character of a vertex operator algebra associated to a four-dimensional superconformal field theory equals the Schur limit of the superconformal index of that theory \cite{Gadde:2011uv}. Furthermore, it was conjectured in \cite{Beem:2017ooy} (as a corollary of the conjecture that the Higgs branch agrees with the associated variety of the associated VOA) that this quantity will always satisfy a finite-order linear modular differential equation. The uniform behavior of the rank-$n$ instanton SCFTs, evidenced in their F-theory description and, at least for $n=1,2$, in their explicitly constructed associated chiral algebras, naturally extends to the modular differential operator annihilating the vacuum character. For $n=1$, this was found to be of second-order with the one free coefficient a quadratic function of the dual Coxeter number \cite{Beem:2017ooy}. Here for $n=2$ we find similarly uniform (twisted) modular differential operators of fourth order. On the basis of evidence coming from the $\mathfrak{d}_4$ case, we further conjecture that for $n=3$ there exist seventh-order twisted modular differential operators, and more generally that for each $n$ there exist uniform modular differential operators annihilating the vacuum characters of the VOAs $\mathcal V_{\gf}^{(n)}$.

The plan of the paper is as follows. In Section \ref{sec:background} we review various known facts about the rank-$n$ instanton SCFTs. In Section \ref{sec:explicit_VOAs} we present explicit vertex operator algebras $\mathcal V_{\gf}^{(2)}$ associated with the rank-two F-theory SCFTs as obtained from the chiral algebra bootstrap approach. In Section \ref{sec:free_fields}, we construct these same VOAs using geometric free field realizations. Section \ref{sec:rank_two_LMDE} is devoted to the fourth-order modular differential operators annihilating the vacuum characters of $\mathcal V_{\gf}^{(2)}$. In Section \ref{sec:four} we briefly discuss the future challenge of developing a more general story for the theories with $n\geqslant3$. We include a variety of useful facts and properties of the Deligne-Cvitanovi\'c series of exceptional Lie algebras in Appendix \ref{app:A}.

\section{Higher rank F-theory SCFTs}
\label{sec:background}

The rank-$n$ F-theory SCFTs describe the low-energy dynamics of a stack of $n$ D3-branes probing a singular fiber of an elliptic $K3$ surface in F-theory on which the dilaton is constant. There are seven such possible singular fibers, and they are typically denoted by $H_0, H_1, H_2, D_4, E_6, E_7$ and $E_8$. The flavor symmetry algebra of the resultant superconformal field theory includes as a simple factor the corresponding Lie algebra $\af_0,\af_1,\af_2,\df_4,\ef_6,\ef_7,\ef_8$, where $\af_0$ represents the trivial Lie algebra. For $n>1$, there is an additional $\suf(2)$ factor in the flavor symmetry. A salient feature of these theories is that their Higgs branches of vacua coincide with the moduli spaces of $n$ centered $\gf$-instantons. (In \cite{Beem:2013sza} it was observed that, from the perspective of the SCFT/VOA correspondence, there is no obstruction to the existence of a theory with flavor symmetry $\gf_2$ or $\mathfrak f_4$ with Higgs branch operators satisfying the relations defining the corresponding one-instanton moduli space. We will see that for the purposes of this paper, the cases $\gf_2$ and $\mathfrak f_4$ continue to be well-behaved.\footnote{It is still unclear if these VOAs arise in connection with actual physical SCFTs. Some arguments against in the case of $\mathfrak f_4$ were presented in \cite{Shimizu:2017kzs}.}) Altogether, the rank-one theories with the property that their Higgs branches are one-instanton moduli spaces are labeled by an algebra belonging to the Deligne-Cvitanovi\'c series of exceptional Lie algebras \cite{Deligne,Cvitanovic:2008zz}:
\begin{equation}
\af_0,\af_1,\af_2,\gf_2,\df_4,\mf{f}_4,\ef_6,\ef_7,\ef_8~.
\end{equation}
In this paper, we aim to study the higher-rank generalizations of these SCFTs, mainly from the viewpoint of the associated vertex operator algebra introduced in \cite{Beem:2013sza}. 

\subsection{Moduli spaces and central charges}
\label{subsec:n_instanton_moduli_cc}

We will begin by recording some useful information about these theories, with an emphasis on the rank-two case.

\subsubsection*{Coulomb branch}

The Coulomb branch chiral ring of a rank-$n$ F-theory SCFT is freely generated by $n$ operators. The $U(1)_r$ charges $r_j, j=1,\ldots, n$ of these generators are integer multiples of the charge of the rank-one generator. In other words
\begin{equation}
\label{eq:CBcharges}
r_j= j\, r~, \qquad \text{with}\qquad r=\frac{\hv+6}{6}~, 
\end{equation}
where $\hv$ denotes the dual Coxeter number of the Lie algebra in question; see Table \ref{tab:deligne} for the relevant values.

In the F-theoretic description, the Coulomb branch corresponds to vacua where the D3-branes are moved away from the singular fiber of an elliptically fibered $K3$ surface.

\renewcommand{\arraystretch}{1.35}
\begin{table}
\centering
\begin{tabular}{|c|c|c|c|c|c|c|}
\hline \hline
~$\gf$~ 		& ~$h^\vee$~	& ~$k_{2d} $~ 			& ~$c_{2d}$~ 		& ~$h_1$~ 				& ~$a_{4d}$~ 				& ~$r$ \\ \hline 
$\gf(h^\vee)$ 	& $h^\vee$ 		& $-\frac{h^\vee\!+6}{6}$ 	& $-2-2h^\vee$ 		& $-\frac{h^\vee}{6}$ 	& $\frac{5+3h^{\vee}}{24}$ 	& $\frac{h^\vee+6}{6}$ \\
\hline 
$\af_0$		& $\frac65$ & $-\frac65$ & $-\frac{22}{5}$ & $-\frac{1}{5}$ & $\frac{43}{120}$ & $\frac{6}{5}$\\\hline 
$\af_1$		& $2$		& $-\frac43$ & $-6$ 	& $-\frac{1}{3}$ & $\frac{11}{24}$ 	& $\frac{4}{3}$\\ \hline
$\af_2$		& $3$		& $-\frac32$ & $-8$		& $-\frac{1}{2}$ & $\frac{7}{12}$ 	& $\frac{3}{2}$ \\\hline
$\gf_2$		& $4$ 		& $-\frac53$ & $-10$	& $-\frac{2}{3}$ & $\frac{17}{24}$ 	& $\frac{5}{3}^{\star}$\\\hline
$\df_4$		& $6$ 		& $-2$		 & $-14$	& $-1$ 			 & $\frac{23}{24}$ 	& 2\\\hline
$\mf{f}_4$	& $9$ 		& $-\frac52$ & $-20$	& $-\frac{3}{2}$ & $\frac{4}{3}$ 	& $\frac{5}{2}^{\star}$\\\hline
$\ef_6$		& $12$ 		& $-3$		 & $-26$	& $-2$ 			 & $\frac{41}{24}$	& $3$\\\hline
$\ef_7$		& $18$ 		& $-4$		 & $-38$	& $-3$ 			 & $\frac{59}{24}$ 	& $4$\\\hline
$\ef_8$		& $30$ 		& $-6$		 & $-62$	& $-5$ 			 & $\frac{95}{24}$ 	& $6$\\\hline
\end{tabular}
\caption{\label{tab:deligne} 
The Deligne-Cvitanovi\'c series of simple Lie algebras, the data of the associated (rank-one) VOAs and the data of their (putative) parent four-dimensional SCFTs. The $\af_0$ entry is a formal member of the list and corresponds to the VOA of the $(2,5)$ Virasoro minimal model, whose four-dimensional parent is the $(A_1,A_2)$ Argyres-Douglas SCFT. As the four-dimensional interpretation of the $\gf_2$ and $\mf{f}_4$ cases is still unclear, the values of $a_{4d}$ and $r$ for these entries are formal/conjectural. In particular even if these theories exist, the values of $r$ may be different if implicit assumptions about their Coulomb branches do not hold.}
\end{table}

\subsubsection*{Higgs branch}

The Higgs branch of the rank-$n$ theory of type $\gf (\neq \af_0)$ is quite a bit more intricate, and can be identified with $\widetilde{\mathcal M}_{\gf}^{(n)}$, the centered $n$-instanton moduli space of $\gf$-instantons on $\mathbb R^4$. This is a hyperk\"ahler manifold with quaternionic dimension
\begin{equation}\label{dim_n_instmodsp}
\dim_{\Hb} \widetilde{\mathcal M}_{\gf}^{(n)} =n\hv -1~.
\end{equation}
This dimension formula can be understood intuitively by considering a configuration of $n$ widely separated one-instanton configurations, each of which has an uncentered moduli space of dimension $\dim_{\Hb}\mathcal M_{\gf}^{(1)} =\hv$, and removing the overall center of mass position. 

As algebraic varieties, the one-instanton moduli spaces of the Deligne-Cvitanovi\'c series of simple Lie algebras have an economical description. Their coordinate rings are generated by adjoint-valued moment maps $\mu_{\gf}$ subject to the Joseph relations \cite{ASENS_1976_4_9_1_1_0}. To present these relations, we first note that one of the defining properties of the Deligne series is the appearance of precisely three real irreducible representations in the decomposition of the symmetric tensor product of two adjoint representations.\footnote{More precisely, the representations are irreducible under $\gf\ltimes \text{Out}(\gf)$, where $\text{Out}(\gf)$ is the group of outer automorphisms of $\gf$.} Following the notations of \cite{cohen1999tensor}, we have
\begin{equation}\label{decompsym2adj}
\mathrm{sym}^2 \mathbf{Adj} = \mathbf{1} \oplus \mathbf{Y_2^\ast} \oplus \mathbf{Y_2}~.
\end{equation}
Here $\mathbf{1} $ denotes the singlet representation, while $\mathbf{Y_k}$ denotes the representation with Dynkin labels $k$ times those of the adjoint representation.\footnote{The $\ast$ defines an involution on the space of representations appearing in various tensor products of the adjoint representation. It has nothing to do with complex conjugation.} Note also that $\af_1$ is slightly degenerate from this point of view, in that the representation $\mathbf{Y_2^\ast}$ is absent. We refer the reader to Appendix \ref{app:A} for additional information about these representations. With this notation established, the Joseph relations state that 
\begin{equation}\label{eqJoseph}
\mu_{\gf}^2\big{|}_{\mathbf{1}}=0~,\qquad \text{and}\qquad \mu_{\gf}^2\big{|}_{\mathbf{Y_2^\ast}}=0~.
\end{equation}

In the rank two theories, the Higgs branch chiral ring has as generators the moment maps $\mu_{\suf(2)}$ and $\mu_{\gf}$, which transform in the $(1,\mathbf{1})$ and $(0,\mathbf{Adj})$ representations of $\suf(2)\times\gf$, respectively, along with an additional multiplet of generators $\omega$ with $SU(2)_R$ charge $R=3/2$ that transforms in the $(\frac12,\mathbf{Adj})$. This collection of generators can, for example, be read off from the two-instanton Hilbert series as computed in \cite{Gaiotto:2012uq,Hanany:2012dm,Keller:2012da,Cremonesi:2014xha}. The Hilbert series also encodes their relations up to numerical coefficients. Decoding that information allows us to write the relations defining the two-instanton moduli space uniformly for all algebras of the Deligne-Cvitanovi\'c series as follows:
\begin{align}
&\text{at $R=2$:} &&\mu_{\suf(2)}^2\big{|}_{(0,\mathbf{1})} = \frac{1}{4}\, \mu_{\gf}^2\big{|}_{(0,\mathbf{1})}~,\label{quadraticrelationHiggs} \displaybreak[0]\\
&\text{at $R=5/2$:} &&\mu_{\gf}\,\omega\big{|}_{(\frac{1}{2},\mathbf{1})}=0~,\label{NonullrelationHiggs}\\
& && \mu_{\gf}\,\omega\big{|}_{(\frac{1}{2},\mathbf{Y_2^\ast})}=0~,\\\label{12Adjcomment}
& && \mu_{\gf}\,\omega\big{|}_{(\frac{1}{2},\mathbf{Adj})} =4\, \mu_{\suf(2)}\,\omega \big{|}_{(\frac{1}{2},\mathbf{Adj})}~,\displaybreak[0]\\
&\text{at $R=3$:} && \omega^2\big{|}_{(1,\mathbf{1})}= -\mu_{\suf(2)}\, \mu_{\gf}^2\big{|}_{(1,\mathbf{1})}~,\label{reln}\\
& && \omega^2\big{|}_{(1,\mathbf{Y_2^\ast})}= -\mu_{\suf(2)}\, \mu_{\gf}^2\big{|}_{(1,\mathbf{Y_2^\ast})}~,\\
& &&  \mu_{\gf}^3 \big{|}_{(0,\mathbf{Adj})} =b_1\,\omega^2\big{|}_{(0,\mathbf{Adj})}~,\\
& && \mu_{\gf}^3 \big{|}_{(0,\mathbf{X_2})}=b_2\,\omega^2\big{|}_{(0,\mathbf{X_2})}~,\\
& && \mu_{\gf}^3 \big{|}_{(0,\mathbf{Y_3^\ast})}=0~,\label{lastHBrelation}
\end{align}
where $b_1$ and $b_2$ are constants that we have not endeavored to fix, though they can be determined straightforwardly using our free-field realizations.\footnote{The normalization of the moment map operators can be understood as in \cite{Beem:2018duj}, and more generally the normalizations used here match the ones used in the free-field realization of the vertex operator algebras $\mathcal V_{\gf}^{(2)}$ presented below in Section \ref{sec:free_fields}. The relation \eqref{12Adjcomment} is simply obtained by taking the Poisson bracket of \eqref{quadraticrelationHiggs} with $\omega$. The numerical constant of proportionality in \eqref{reln} can be taken to define the normalization of $\omega$.}\textsuperscript{,}\footnote{More explicitly, \eqref{12Adjcomment} reads $i f^{A}_{\phantom{A}BC}\,\mu_{\mathfrak{g}}^B \omega_{\alpha}^C=4\, (\mu_{\suf(2)})_{\alpha\beta}\,\omega^A_{\gamma} \epsilon^{\beta \gamma}$.} Indeed, these realizations offer an efficient way to determine the full complement of Higgs branch relations more generally. Here we have made use of the uniform representation content of the decomposition of $\mathrm{sym}^3 \mathbf{Adj}$,
\begin{equation}\label{decompsym3adj}
\mathrm{sym}^3 \mathbf{Adj} = \mathbf{Adj} + \mathbf{X_2} + \mathbf{A} + \mathbf{Y_3} + \mathbf{Y_3^\ast}~.
\end{equation}
Again, see Appendix \ref{app:A} for representation-theoretic details.

For still higher-rank theories, the Higgs branch chiral ring generators have been conjectured to have the following quantum numbers \cite{Cremonesi:2014xha}  
\begin{align}
\label{Higgsgen1}\big(\tfrac{\ell}{2},\mathbf{Adj} \big)_{R\,=\,\frac{\ell+2}{2}}&\,,\qquad\,\ell\,=0,1,\dots,n-1~,\\
\label{Higgsgen2}\big(\tfrac{m+1}{2},\mathbf{1} \big)_{R\,=\,\frac{m+1}{2}}&\,,\qquad m=1,2,\dots,n-1~.
\end{align}
The symmetry properties of the chiral ring relations can also in principle be extracted from the Hilbert series computed in \cite{Cremonesi:2014xha}.
 
The rank-$n$ theories of type $\mathfrak{a}_0$ are exceptional in that they possess only $\mathfrak{su}(2)$ flavor symmetry. The Higgs branches for these theories coincides with the Higgs branches of $\mathcal{N}=4$ SYM with gauge algebra $\mathfrak{su}(n)$, namely $(\mathbb{C}^2)^{n-1}/S_n$. See, \emph{e.g.}, \cite{Bonetti:2018fqz} for additional discussion of these Higgs branches. As opposed to $\mathcal{N}=4$ SYM, where at a generic point on the Higgs branch the spectrum consists of free vector multiplets, in the case of the rank-$n$ $\mathfrak{a}_0$ SCFTs, $n$ copies of the rank-one theory survive. The latter has a trivial Higgs branch.

\subsubsection*{Central charges}

Central charges of higher-rank F-theory SCFTs were first computed in \cite{Aharony:2007dj} using holographic methods. For the rank-$n$ theory of type $\gf$, the $a$ and $c$ Weyl anomaly coefficients and the $\suf(2)$ and $\gf$ flavor central charges were computed to take the following values:
\begin{align}
a_{4d} 				&= \frac{1}{24}(-1+6n^2+n(2+n)\hv)~,\label{a-anomaly}\\
c_{4d} 				&= \frac{1}{24}(-2+6n^2+n(3+n)\hv)~,\\
k_{4d}^{\suf(2)}	&= \frac{1}{6}(n-1)(6+n(6+\hv))~,\\
k_{4d}^{\gf} 		&= \frac{n}{3}(6+\hv)~.
\end{align}
Let us pause to make a few observations about these results. First of all, for $\mathfrak{g}\neq\mathfrak{a}_0$ these theories have no residual massless degrees of freedom in generic Higgs branch vacua aside from the hypermultiplets that parameterize the Higgs branch. As a result, the quaternionic dimension of the Higgs branch $\MM_H$ is recovered from the difference of the $a$ and $c$ Weyl anomaly coefficients according to the relation
\begin{equation}\label{eq:dim_higgs_a-c}
\dim_{\Hb} \MM_{H} = -24(a_{4d}-c_{4d}) = n\hv-1~,
\end{equation}
which agrees with the dimension reported in \eqref{dim_n_instmodsp}. Alternatively, in the $\mathfrak{a}_0$ theory, we have
\begin{equation}
-24(a_{4d}^{(n)}-c_{4d}^{(n)}) = \frac{6n-5}{5} = \dim_{\Hb}\MM_H-24n\left(a_{4d}^{(1)}-c_{4d}^{(1)}\right)~,
\end{equation}
which follows from anomaly matching given that the theory on the Higgs branch includes $n$ copies of the rank-one $\mathfrak{a}_0$ SCFT.

Additionally, one can verify that the Shapere-Tachikawa relation between Weyl anomaly coefficients and Coulomb branch data holds \cite{Shapere:2008zf}. Indeed, using the charges in \eqref{eq:CBcharges}, one checks
\begin{equation}
2a_{4d} - c_{4d} = \frac{1}{4} \sum_{j=1}^n (2 r_j - 1) = \frac{n}{24}(\hv + (6+\hv)n)~.
\end{equation}
The previous two relations for $a_{4d}$ and $c_{4d}$ could in principle have been used to find these anomaly coefficients directly from Higgs and Coulomb branch data. Also note that for $n=1$ the flavor central charge of the $\suf(2)$ symmetry algebra is zero, which indicates its absence for rank-one theories. Finally, it is of note that the flavor central charges of the $\gf$ currents is linear in $n$. In Tables \ref{tab:deligne} and \ref{tab:rank2_deligne} we display these and other pieces of discrete numerical data for the rank-one and rank-two theories. We give the data in terms of the rescaled quantum numbers that are directly related to properties of the vertex algebras associated to the four-dimensional SCFTs,
\begin{equation}\label{eq:2d_central_charges}
k_{2d} = -\frac{1}{2} k_{4d}~,\qquad c_{2d} = -12 c~. 
\end{equation}
The former gives the level of the respective affine current subalgebras of the associated VOA, which arise as an enhancement of the four-dimensional flavor symmetries, and the latter is the VOA's Virasoro central charge.

\renewcommand{\arraystretch}{1.35}
\begin{table}
\centering
\begin{tabular}{|c|c|c|c|c|c|c|c|}
\hline \hline
~$\gf$~ 	& ~$h^\vee$~& ~$k^{\gf}_{2d}$~ & ~$k^{\suf(2)}_{2d}$~ & ~$c_{2d}$~ & ~$\tilde{h}_{\rm min}$~ & ~$a_{4d}$~ & ~$r_1$ \\ 
\hline 
$\gf(h^\vee)$ & $h^\vee$ & $-\frac{h^\vee\!+6}{3}$ &$-\frac{h^\vee\!+9}{6}$ & $-11-5\hv$ & $-\frac{9+9h^\vee}{24}$ & $\frac{23+8h^\vee}{24}$ & $\frac{h^\vee+6}{6}$\\
\hline 
$\af_0$		& $\frac65$ & ~--- & $-\frac{17}{10}$ &$-17$ & $-\frac{33}{40}$ & $\frac{163}{120}$ & $\frac{6}{5}$\\
\hline 
$\af_1$		& $2$ 		& $-\frac83$ & $-\frac{11}{6}$ &$-21$ 	& $-\frac{9}{8}$ & $\frac{13}{8}$ 	& $\frac{4}{3}$\\ 
\hline
$\af_2$		& $3$		& $-3$ 		 & $-2$ & $-26$	& $-\frac{3}{2}$ & $\frac{47}{24}$ & $\frac{3}{2}$ \\
\hline
$\gf_2$		& $4$ 		& $-\frac{10}{3}$ & $-\frac{13}{6}$ & $-31$	& $-\frac{15}{8}$ & $\frac{55}{24}$ 	& $\frac{5}{3}^{\star}$\\
\hline
$\df_4$		& $6$ 		& $-4$		 & $-\frac{5}{2}$ &$-41$	& $-\frac{21}{8}$ 	& $\frac{71}{24}$ 	& 2\\
\hline
$\mf{f}_4$	& $9$ 		& $-5$ 		 & $-3$ &$-56$	& $-\frac{15}{4}$ & $\frac{95}{24}$ 	& $\frac{5}{2}^{\star}$\\
\hline
$\ef_6$		& $12$ 		& $-6$		 & $-\frac{7}{2}$ &$-71$	& $-\frac{39}{8}$ 			 & $\frac{119}{24}$	& $3$\\
\hline
$\ef_7$		& $18$ 		& $-8$		 & $-\frac{9}{2}$ & $-101$	& $-\frac{57}{8}$ 			 & $\frac{167}{24}$ 	& $4$\\
\hline
$\ef_8$		& $30$ 		& $-12$		 & $-\frac{13}{2}$ & $-161$	& $-\frac{93}{8}$ 			 & $\frac{263}{24}$ 	& $6$\\
\hline
\end{tabular}
\caption{\label{tab:rank2_deligne} 
Data for the rank-two Deligne-Cvitanovi\'c VOAs and their parent four-dimensional SCFTs. As for rank one, the four-dimensional interpretation of the $\gf_2$ and $\mf{f}_4$ cases is unresolved.}
\end{table}

In light of the various unitarity bounds derived in \cite{Beem:2013sza,Lemos:2015orc,Beem:2017ooy,Beem:2018duj}, one should observe one additional fact about the rank-two theories.\footnote{The saturation of unitarity bounds of rank-one F-theory SCFTs has already been analyzed in great detail in the aforementioned papers.} For those theories, the sum of the Sugawara central charges of the $\suf(2)$ and $\gf$ current algebras matches the total central charge,
\begin{equation}\label{sugawaracondition}
\text{for rank-2 theories:} \qquad c_{2d} = c_{\mathrm{Sug}}^{\suf(2)} + c_{\mathrm{Sug}}^{\gf} = \frac{k_{2d}^{\suf(2)} \dim \suf(2)}{k_{2d}^{\suf(2)} + h^\vee_{\suf(2)}} +  \frac{k_{2d}^{\gf} \dim \gf}{k_{2d}^{\gf} + h^\vee_\gf}~.
\end{equation}
In the four-dimensional physics of the rank-two theories, this equality is reflected in the Higgs branch relation \eqref{quadraticrelationHiggs}.\footnote{This Higgs branch chiral ring relation is actually a necessary consequence of the Sugawara value of the central charge \cite{Beem:2018duj}.} In the associated VOA, this implies the absence of an independent stress energy tensor as a strong VOA generator. Instead, this role is taken over by the total Sugawara stress tensor. 

An exceptional cases arises for $\gf=\af_2$, where both the $\suf(2)$ and $\suf(3)$ current algebras are at their respective critical levels, \emph{i.e.}, $k_{2d} = -\hv$. Consequently for this VOA the Sugawara construction fails to furnish a normalizable stress tensor for both factors, and a separate stress tensor will be a strong generator of the associated VOA. The Higgs branch relation \eqref{quadraticrelationHiggs} in this case follows from a second unitarity argument from \cite{Beem:2018duj}.

\subsection{Class \texorpdfstring{$\SS$}{S} realizations}
\label{subsec:class_S_realizations}

The higher-rank theories that are not of Argyres-Douglas type can be realized within class $\SS$ using only regular punctures \cite{Benini:2009gi,Benini:2010uu,Moore:2011ee}. What's more, the higher-rank $\df_4$ theories admit conventional Lagrangian descriptions. In this subsection, we briefly recall these realizations. 

Denoting by $\TT[\gf,\CC_{g,s},\{\Lambda_i\}]$ the class $\SS$ theory obtained by (partially) twisted compactification of the $(2,0)$ theory of type $\gf$ on a Riemann surface $\CC_{g,s}$ of genus $g$ with $s$ punctures, with choices $\Lambda_i: \suf(2) \hookrightarrow \gf$ of embeddings of $\suf(2)$ into $\gf$ for each puncture, the rank-$n$ theories of type $\df_4,\ef_6,\ef_7,\ef_8$ are realized as
\begin{equation}
\begin{aligned}\label{eq:class_S_realizations}
& \text{rank-$n$ $\df_4$ theory}\qquad &\longleftrightarrow& \qquad \TT[\af_{2n-1},\CC_{0,4},\{[n^2],[n^2],[n^2],[n^2]\}]~,\\
& \text{rank-$n$ $\ef_6$ theory} \qquad &\longleftrightarrow& \qquad\TT[\af_{3n-1},\CC_{0,3},\{[n^3],[n^3],[n^3]\}]~,\\
& \text{rank-$n$ $\ef_7$ theory} \qquad &\longleftrightarrow& \qquad\TT[\af_{4n-1},\CC_{0,3},\{[n^4],[n^4],[(2n)^2]\}]~,\\
& \text{rank-$n$ $\ef_8$ theory} \qquad &\longleftrightarrow& \qquad\TT[\af_{6n-1},\CC_{0,3},\{[n^6],[(2n)^3],[(3n)^2]\}]~,\\
\end{aligned}
\end{equation}
where the embeddings $\Lambda_i$ are represented by a partition of the rank of the relevant $\af$-type algebra plus one.\footnote{Exponents always denote repeated entries in the partition.} In the ``good-bad-ugly'' trichotomy introduced in \cite{Gaiotto:2012uq}, these theories are all ``bad'', which in particular means that the prescription of \cite{Gadde:2011uv} to compute their superconformal indices diverges.\footnote{This prescription was derived in \cite{Gaiotto:2012xa} by demanding that the superconformal index be (generalized) S-duality invariant. However, ``bad'' theories do not participate in the S-duality web, as they do not admit any exactly marginal gaugings.} In \cite{Gaiotto:2012uq} a proposal was put forward for ``ugly'' theories (\emph{i.e.}, theories containing decoupled free hypermultiplets) whose interacting sectors are again precisely these higher-rank theories. Concretely,
\begin{equation}
\begin{aligned}
& \text{rank-$n$ $\df_4$  $\oplus$ 1 free HM}\quad &\longleftrightarrow& \qquad \TT[\af_{2n-1},\CC_{0,4},\{[n^2],[n^2],[n^2],[n,n-1,1]\}]~,\\
& \text{rank-$n$ $\ef_6$  $\oplus$ 1 free HM} \quad &\longleftrightarrow& \qquad\TT[\af_{3n-1},\CC_{0,3},\{[n^3],[n^3],[n^2,n-1,1]\}]~,\\
& \text{rank-$n$ $\ef_7$  $\oplus$ 1 free HM} \quad &\longleftrightarrow& \qquad\TT[\af_{4n-1},\CC_{0,3},\{[n^4],[n^3,n-1,1],[(2n)^2]\}]~,\\
& \text{rank-$n$ $\ef_8$  $\oplus$ 1 free HM} \quad &\longleftrightarrow& \qquad\TT[\af_{6n-1},\CC_{0,3},\{[n^5,n-1,1],[(2n)^3],[(3n)^2]\}]~.\\
\end{aligned}
\end{equation}
Finally, we note that the rank-$n$ $\df_4$ theories admit Lagrangian descriptions as $\mathfrak{usp}(2n)$ gauge theories with four hypermultiplets transforming in the fundamental representation of $\mathfrak{usp}(2n)$ and one hypermultiplet in the antisymmetric representation \cite{Douglas:1996js}.

\section{Explicit VOA constructions}
\label{sec:explicit_VOAs}

We now turn to the main task of this paper, which is to construct explicitly the rank-two associated VOAs $\mathcal{V}^{(2)}_{\gf}$. For those cases which are not Argyres-Douglas type, one has in principle algorithmic constructions coming from the relevant class $\mathcal{S}$ or Lagrangian descriptions of these theories. In particular, the class $\mathcal{S}$ theories given in \eqref{eq:class_S_realizations} can be used to give a definition of the associated variety in terms of a BRST reduction of several equivariant affine $\mathcal{W}$-algebras as described in \cite{Arakawa:2018egx}, while the Lagrangian realization of the $\df_4$ theories give a BRST description as described in \cite{Beem:2013sza}. However, both of these approaches present severe technical challenges in the computation of the relevant BRST cohomologies.

Our strategy instead is to make a motivated Ansatz for the set of strong generators of the VOA and demand that they indeed generate a consistent, nontrivial chiral algebra. In practice, we write down the most general expressions, compatible with the global symmetries for the singular terms in the operator product expansions of strong generators in terms of a number of undetermined, numerical coefficients, and impose that the Jacobi identities hold true. The resulting, typically quadratic, equations for the coefficients admit a solution if the Ansatz for the generators correctly describes a (sub)algebra of the sought-after VOA.\footnote{Note that the Jacobi identities are not necessarily zero on the nose, but should hold only up to null fields. A convenient strategy to impose the correct non-null constraints is to set to zero all two-point functions of the right-hand side of the Jacobi identities with a basis of fields of the appropriate quantum numbers.} Such a strategy has been pursued in the context of chiral algebras associated with four-dimensional $\mathcal N=2$ SCFTs in \cite{Beem:2014rza,Lemos:2014lua}, and has been successfully applied in various instances since.\footnote{In our computations we have utilized the \textsc{Mathematica} package developed in \cite{Thielemans:1991uw}.}

It was proven in \cite{Beem:2013sza} that the generators of the Higgs branch chiral ring of a four-dimensional $\mathcal{N}=2$ SCFT necessarily give rise to a (not necessarily strict) subset of the strong generators of the associated VOA.\footnote{In fact, a more general statement is that generators of the Hall-Littlewood chiral ring descend to strong generators of the associated VOA. For the theories under consideration, however, the Hall-Littlewood chiral ring and Higgs branch chiral ring coincide.} For all rank-two VOAs barring $H_0$ and $H_2$, we will therefore make the minimal Ansatz that these constitute the full set of strong generators.\footnote{One could contemplate the presence of additional generators of non-Higgs branch type. However, any such additional generators will be strongly constrained by the fact that the stress tensor takes Sugawara form, and therefore the dimensions of any additional affine Kac-Moody primary will be determined by its $\suf(2)\times\gf$ representation. Indeed, one can show on this account that for $\gf\neq\df_4$, any additional strong generators will not appear in the OPEs of the strong generators of Higgs branch type. In other words, we are guaranteed to find a consistent subalgebra.} As reviewed in section \ref{sec:background}, the Higgs branch chiral ring of these rank-two theories is generated by moment map operators $(\mu_{\suf(2)})_{(\alpha\beta)}$ and $\mu^A_{\gf}$, transforming in the adjoint representation of $\suf(2)$ and $\gf$ respectively, and an additional generator $\omega_{\alpha}^{A}$ transforming in the $(\frac12,{\bf Adj})$ representation of $\suf(2)\times\gf$. Here we have traded the adjoint index of $\suf(2)$ with a symmetrized pair of fundamental indices. A standard entry of the SCFT/VOA correspondence states that the moment map operators give rise to affine currents in the associated VOA. Their levels were given as a function of the dual Coxeter number in the previous section:
\begin{equation}
k_{2d}^{\suf(2)} = -\frac{\hv+9}{6}~,\quad k_{2d}^{\gf} = -\frac{\hv+6}{3}~. 
\end{equation}
See also Table \ref{tab:rank2_deligne}. Table \ref{tab:rank2_deligne_VOAgenerators} summarizes our notations and the quantum numbers for the strong VOA generators that we are postulating. We have exploited the equality between the conformal weight of strong generators associated to Higgs branch chiral ring generators and the $SU(2)_R$ charge of those Higgs branch chiral ring operators.
\renewcommand{\arraystretch}{1.35}
\begin{table}
\centering
\begin{tabular}{|c|c|c|c|}
\hline \hline
~$\mathcal O$~ 	& ~$\chi[\mathcal O]$~& ~$h_{\mathcal O}$~ & ~$\suf(2)\times\gf$ representation~ \\ 
\hline 
~$(\mu_{\suf(2)})_{(\alpha\beta)}$~ 	& ~$j_{(\alpha\beta)}$~& ~$1$~ & ~$(1,\mathbf 0)$~ \\ 
\hline
~$\mu^{A}_{\gf}$~ 	& ~$\JJ^{A}$~& ~$1$~ & ~$(0,\mathbf{Adj})$~ \\ 
\hline
~$\omega_\alpha^{A}$~ 	& ~$\WW_\alpha^{A}$~& ~$\frac{3}{2}$~ & ~$(\frac12,\mathbf{Adj})$~ \\ 
\hline
\end{tabular}
\caption{\label{tab:rank2_deligne_VOAgenerators} 
Generators of the rank-two vertex operator algebras. $H_2$ additionally possesses an independent stress tensor $\TT$, while for $H_0$ $\gf$ is trivial and the weight $3/2$ generator is replaced by a weight $5/2$ doublet generator $w_{\alpha}$.}
\end{table}%

Some special consideration is necessary for the $H_0$ and $H_2$ theories. For $H_2$, the VOA stress tensor can no longer be furnished by the Sugawara construction due to the criticality of the current algebra levels, which obstructs the construction of a normalizable stress tensor. Consequently, an additional, independent Virasoro stress tensor will have to be included as a strong generator of $\mathcal{V}^{(2)}_{\af_2}$. On the other hand, for $H_0$ the VOA associated to Higgs branch generators is just the affine Kac-Moody VOA $V_{-\frac{17}{10}}(\slf(2))$, but there is the possibility that this algebra should be extended by additional (non-Higgs) strong generators. 

Indeed, one can see that such extra strong generators must be included on the basis of consistency under Higgsing. If we consider a Higgs branch vacuum of the four-dimensional SCFT where the $\suf(2)$ moment map operator acquires a nilpotent vacuum expectation value, then there should be two copies of the rank-one $H_0$ theory remaining at low energies. At the level of the VOA, this Higgsing can be accomplished by quantum Drinfel'd-Sokolov (DS) reduction \cite{Beem:2014rza}. However, if the rank-two VOA is generated only by affine currents, then DS reduction will give a single copy of the Virasoro VOA ${\rm Vir}_{3,10}$ with central charge $c=-\frac{44}{5}$. We thus must extend the affine current algebra. From $k^{\suf(2)}_{2d}+2 = \frac{3}{10}$, we learn that the affine level is admissible, as $p'=3$ and $p=10$ are coprime and $p'>1$. An admissible current algebra can only be extended by AKM primaries of spins $\frac{1}{2},1,\ldots,\frac{p'-2}{2}$, see \cite{Adamovic}. Hence our only option is an extension by a spin $\frac{1}{2}$ AKM primary. The dimension of this primary follows from the standard formula $h=j(j+1)/(k^{\suf(2)}_{2d}+2)$ and gives $h=\frac{5}{2}$. Thus we are led to introduce an extra $\suf(2)$ doublet of strong generators of weight $\frac{5}{2}$, which in four dimensions should arise from a Schur operator in a $\hat{\mathcal{C}}_{1/2,(0,0)}$ multiplet.

It remains to describe the singular OPEs of our strong generators. The $\suf(2)\times\gf$ affine current algebras take a standard form,\footnote{Our conventions for the epsilon tensor are $\epsilon^{12}=-\epsilon^{21} = -\epsilon_{12}=\epsilon_{21}=1$.}
\begin{align}\label{jjOPE}
j_{\alpha\beta}(z)~j_{\gamma\delta}(w) &~\sim~ \frac{k_{2d}^{\suf(2)}\ \epsilon_{\delta(\alpha}\epsilon_{\beta)\gamma}}{(z-w)^2} ~+~ \frac{2 \epsilon_{(\alpha(\gamma}\ j_{\beta)\delta)}(w)}{z-w}~,\\\label{JJOPE}
\JJ^{A}(z)~\JJ^{B}(w) &~\sim~ \frac{k_{2d}^\gf \ \kappa^{AB}}{(z-w)^2} ~+~ \frac{i f^{AB}_{\phantom{AB}C}\  \JJ^C(w)}{z-w}~,
\end{align}
while $j_{\alpha\beta}$ and $\mathcal{J}$ have regular OPEs with one another. Parentheses around indices denote symmetrization with weight one and $k_{2d}^{\suf(2)}$ and $k_{2d}^{\gf}$ are given in Table \ref{tab:rank2_deligne_VOAgenerators}. The transformation properties of the additional $\WW_\alpha^A$ generator under the global $\suf(2)\times\gf$ symmetry completely determine its operator product expansion with the currents,
\begin{equation}\label{jWOPEs}
j_{\alpha\beta}(z)~\WW_\gamma^A(w) ~\sim~ \frac{\epsilon_{(\alpha|\gamma|}\ \WW_{\beta)}^{A}(w)}{z-w}~,\qquad \JJ^A(z)~\WW_\alpha^B(w)~\sim~\frac{if^{AB}_{\phantom{AB}C}\ \WW^C_\alpha(w)}{z-w}~.
\end{equation}
For the $H_2$ theory, the independent stress tensor has self-OPE and OPE with the other generators taking their canonical form. For $H_0$, we omit the $\gf$ affine currents. The OPE of the $\suf(2)$ currents with $w_{\alpha}$ is analogous to the first expression in \eqref{jWOPEs}. Our remaining (and primary) task is to find the self-OPE of $\WW$ (or $w$). We will separately consider $H_0$, which possesses certain special features, and present all other instances in one fell swoop.

\subsection{\texorpdfstring{$H_0$}{H0} theory and an \texorpdfstring{$\suf(2)$}{su(2)} extension}
\label{subsec:H0_bootstrap}

To complete the description of the rank-two $H_0$ VOA we need the self-OPE of the additional generator $w_\alpha$. The most general expression compatible with $\mathfrak{su}(2)$ covariance takes the form
\begin{align}
\label{wwOPEH0}
w_{\alpha}(z)~w_{\beta}(w)  &~\sim~ \frac{c_1\ \epsilon_{\alpha\beta}}{(z-w)^5} ~+~ \frac{c_2\ j_{\alpha\beta}(w)}{(z-w)^4} ~+~ \frac{c_3\  \partial j_{\alpha\beta}(w) + c_4\  \epsilon_{\alpha\beta}\ (jj)(w)}{(z-w)^3} \nn \\
&~+~ \frac{1}{(z-w)^2}\Big( c_5 \ \partial^2 j_{\alpha\beta}(w) + c_6\ \epsilon_{\alpha\beta} (j\partial j)(w) \nn \\
&\qquad \qquad\qquad  + c_7\ (j_{(\alpha|\gamma|}\partial j_{\beta)\delta})(w) \ \epsilon^{\gamma\delta} + c_8\ ((jj)j_{\alpha\beta})(w) \Big) \nn\\
&~+~ \frac{1}{z-w}\Big( c_9\ \partial^3j_{\alpha\beta}(w) + c_{10}\ \epsilon_{\alpha\beta}\ (\partial j \partial j)(w) + c_{11}\ \epsilon_{\alpha\beta}\ (j \partial^2 j)(w) \nn\\[3pt]
&\qquad \qquad\qquad +c_{12}\ (j_{(\alpha|\gamma|}\partial^2 j_{\beta)\delta})(w) \ \epsilon^{\gamma\delta} + c_{13}\ ((jj)\partial j_{\alpha\beta})(w) \nn \\[3pt]
&\qquad \qquad\qquad + c_{14}\ (j_{(\alpha|\gamma|}j_{\beta)\delta}\partial j_{\zeta\eta})(w)\ \epsilon^{\gamma\zeta}\epsilon^{\delta\eta} + c_{15} \ \epsilon_{\alpha\beta}\ ((jj)(jj))(w)\Big)~.
\end{align}
We have adopted the short-hand notation $(VV)$, for any operator $V$ transforming in the triplet of $\suf(2)$, to denote the contraction $(VV) = V_{\gamma\delta}\ V_{\zeta\eta}\ \epsilon^{\gamma\zeta}\epsilon^{\delta\eta}$. Here and throughout this section, composite operators are defined by nested normal ordering: $X_1 X_2 \ldots X_{n-1} X_n \colonequals (X_1 (X_2(\ldots (X_{n-1} X_n)\ldots)))$. Normal ordering brackets take precedence over brackets introduced to delineate group theory contractions. 

Jacobi identities uniquely determine the coefficients $c_{i}$ up to a choice of normalization for the generator $w_\alpha$, which we fix by choosing $c_{1}=1$. The resulting structure constants are then as follows:
\begin{alignat}{5}\label{csforH0}
&c_1 = 1~,&& \qquad c_2 = \frac{10}{17}~,&& \qquad c_3 = \frac{5}{17}~,&& \qquad c_4 =\frac{25}{51}~,&& \qquad c_5=-\frac{545}{561}~,\nn\\
&c_6 = \frac{25}{51}~,&&\qquad  c_7=-\frac{1150}{561}~,&& \qquad c_8=\frac{250}{561}~,&& \qquad c_9 = -\frac{35}{132}~,&& \qquad c_{10}=\frac{175}{748}~,\\
&c_{11}=\frac{50}{561}~,&& \qquad c_{12}=-\frac{325}{561}~,&& \qquad c_{13}=\frac{250}{561}~,&& \qquad c_{14}=\frac{250}{561}~,&& \qquad c_{15} = \frac{625}{3366}~.\nn
\end{alignat}
It may be worth noting that the Jacobi identities can be satisfied (up to null states, and with different values for the coefficients $c_i$) for two additional $\suf(2)$ levels other than $k_{2d}^{\suf(2)}=-\frac{17}{10}$, namely $k_{2d}^{\suf(2)}=-\frac{5}{2}$ and $k_{2d}^{\suf(2)}=-\frac{7}{4}$.

\subsection{A universal expression for rank-two VOAs \texorpdfstring{$\mathcal{V}^{(2)}_{\gf}$}{V(2)(g)}}
\label{subsec:general_rank_two_VOA}

In terms of the universal decomposition of the symmetric product of two adjoint representations given in \eqref{decompsym2adj}, we can construct a general expression for the self-OPE of $\WW$ for the rank-two VOAs other than $H_0$ as follows,
\begin{align}
\label{WWOPEs}
\WW_\alpha^A(z)~\WW_\beta^B(w) ~\sim~ &\frac{c_1\,\epsilon_{\alpha\beta}\,\kappa^{AB}}{(z-w)^3} ~+~ \frac{i c_2\,\epsilon_{\alpha\beta}\,f^{AB}_{\phantom{AB}C}\,\JJ^C(w) + c_3\,\kappa^{AB}\,j_{\alpha\beta}(w)}{(z-w)^2}\nn\\[4pt]
~+~& \frac{1}{z-w}\Big(ic_4\,\epsilon_{\alpha\beta}\,f^{AB}_{\phantom{AB}C}\,\partial\JJ^C(w) + c_5\,\kappa^{AB} \,\partial j_{\alpha\beta}(w)  \nn\\[4pt]
&\qquad\qquad + c_6\,\kappa^{AB}\,\epsilon_{\alpha\beta} \,(jj)(w)+ i\,c_7f^{AB}_{\phantom{AB}C}\,(j_{\alpha\beta}\JJ^C)(w)\\[5pt]
&\qquad\qquad + \epsilon_{\alpha\beta}\big( c_8\,\mathds{Y}_2^{(AB)}(w)+ c_9\,\mathds{1}^{(AB)}(w) + c_{10}\,\mathds{Y}_2^{*(AB)}(w)\big)  \Big)\nn\\[4pt]
~+~& \frac{\epsilon_{\alpha\beta}\,\kappa^{AB} \big(-\frac{1}{4} + (\hv-3) b\big)}{z-w} \bigg( \mathcal T(w) - \frac{\kappa_{CD}(\mathcal J^C \mathcal J^D)(w)}{2(k_{2d}^{\gf} + \hv)} + \frac{(jj)(w)}{2(k_{2d}^{\suf(2)} + 2)}  \bigg)~.\nn
\end{align}
The short-hand notation $(jj)$ continues to denote the contraction $(jj) = j_{\gamma\delta}\ j_{\zeta\eta}\ \epsilon^{\gamma\zeta}\epsilon^{\delta\eta}$, and furthermore we have introduced the notations $\mathds{1}^{(AB)}$, $\mathds Y_2^{(AB)}$, and $\mathds{Y}_2^{*(AB)}$ to represent the projections of the product of two (adjoint) currents $\mathcal J$ onto the respective representations, \emph{i.e.}, $\mathds{1}^{(AB)} = (P_{\mathbf 1})_{CD}^{AB}\ \JJ^C\JJ^D = \frac{\kappa^{AB}\kappa_{CD}\mathcal J^C\mathcal J^D}{\dim\mathbf{Adj}}$, and similarly for $\mathds{Y}_2^{(AB)}$, and $\mathds{Y}_2^{*(AB)}$. We provide the precise expressions for the relevant projection operators in Appendix \ref{app:A}. Up to a choice of normalization of the generator $\WW$, which we set by choosing $c_1=1$, the coefficients $c_i$ are uniquely fixed in terms of the dual Coxeter number $\hv$ as
\small
\begin{align}
&c_1 =1~, \qquad c_2 = -\frac{3}{6+\hv}~, \qquad c_3 = \frac{6}{9+\hv}~, \qquad c_4 = -\frac{3}{2(6+\hv)}~, \qquad  c_5 =  \frac{3}{9+\hv}~, \nn\\
&c_6 = -\frac{3}{4(\hv-3)} + \frac{3}{4(\hv+9)} ~,\qquad c_7 = -\frac{18}{(6+\hv)(9+\hv)}~,\qquad c_8 = -\frac{9}{(6+\hv)(9+\hv)}~, \nn\\
&\frac{c_9}{\dim\mathbf{Adj}} = -\frac{3}{16(\hv-3)} + \frac{3(2+5\hv)}{16(1+\hv)(-6+5\hv)} ~, \quad  c_{10} = \frac{3}{6+\hv}~. \label{coefficientsDeligne}
\end{align}
\normalsize
Note that in \eqref{WWOPEs} we included a stress tensor $\mathcal T$. As explained above, for $\gf\neq \af_2$, \emph{i.e.}, $\hv \neq 3$, it is actually given by the Sugawara construction, and so the last line represents a null operator that can be added with arbitrary coefficient (controlled by the parameter $b$). In the limit of $\hv\rightarrow 3$, however, $\mathcal{T}$ is truly an independent strong generator. We can see from \eqref{coefficientsDeligne} that this limit can in fact be taken smoothly as the poles in $\hv-3$ are cancelled. In this limit, the arbitrary coefficient $b$ appears in front of a null operator of the $H_2$ VOA. In \eqref{coefficientsDeligne} we have divided $c_9$ by the dimension of adjoint representation of $\gf$ to directly reflect the coefficient of $\kappa_{CD}\mathcal J^C\mathcal J^D$. This is a helpful way of writing things in view of taking the limit for $\hv\rightarrow 3$. 

The rank-two VOA associated to $H_1$ can be recovered by specializing to $\hv=2$ and omitting the operator $\mathds{Y}_2^{*(AB)}$ on the right hand side of \eqref{WWOPEs} -- the corresponding projector vanishes identically for $\suf(2)$. Incidentally, the $H_2$ theory has already been studied in detail in the literature from the perspective of the associated VOA in \cite{Buican:2017fiq}, though it was not identified as being the rank-two instanton SCFT in that paper, but rather was recovered in the strong coupling limit of a conformal gauging of Argyres-Douglas SCFTs and given the moniker $\TT_X$.

\section{Free field realizations}
\label{sec:free_fields}

Before pursuing any extensive analysis of the VOAs $\mathcal{V}_{\fg}^{(2)}$ constructed in the previous section, it is worth investigating whether these algebras admit free field realizations in the style of \cite{Beem:2019tfp}. Such a realization has the potential to simplify the analysis of singular vectors in the vacuum module, as well as providing a canonical proposal for the four-dimensional $R$-filtration of these VOAs \cite{Beem:2017ooy}. 

According to the template introduced in that paper, we expect a realization in terms of $\dim_{\Cb}\MM_H = 2(2 h^\vee-1)$ chiral bosons associated to a lattice of signature $(2h^\vee-1,2h^\vee-1)$, whose lattice momenta are restricted to an isotropic sublattice; the construction should also reflect the algebraic structure of the Higgs branch/associated variety.

\subsection{Realizations for \texorpdfstring{$\gf\neq\mathfrak{a}_0$}{gf not a0} from intermediate Higgsing}
\label{subsec:general_case_FFR}

It turns out that a more efficient approach will be to develop an intermediate construction associated not with the generic locus of the Higgs branch, but with the singular stratum where the $\gf$ symmetry is unbroken. (For the special case of the $H_0$ theory, this is indeed the generic locus.) The idea is that we can use lattice bosons to model the geometry of the singular locus, but the ``free field realization'' should be further decorated with the VOAs associated to the residual degrees of freedom on that locus. In the present case, the residual theory on the singular locus in question is two copies of the rank-one SCFT, so our free field realizations will include two copies of the rank-one VOA $\mathcal{V}_{\fg}^{(1)}$ as basic building blocks. Given the free-field constructions of the rank-one VOAs in \cite{Beem:2019tfp}, our final result could then be further expressed as an honest free field realization in terms of only chiral lattice bosons and symplectic bosons.

\subsubsection{Big open sets in the two-instanton moduli spaces}
\label{subsubsec:finite_geometry}

On the locus of $\MM_H$ where the $\gf$ symmetry is unbroken, the moment map $\mu_{\gf}$ and all the chiral ring generators charged under $\gf$, namely $\omega^A_{\alpha}$, vanish, 
\begin{equation}
\langle\mu^A_{\gf}\rangle=\langle \omega^A_{\alpha} \rangle=0~.
\end{equation}
This locus is then parameterized by the $\suf(2)$ moment map, $\mu_{\suf(2)}$, subject to the relation \eqref{quadraticrelationHiggs}, which implies that
\begin{equation}
\mu_{\suf(2)}^2\big{|}_{(0,\mathbf{1})}=0~.
\end{equation}
In other words, the singular locus in question is a copy of $\overline{\mathbb{O}_{\min}(\slf(2))}\cong\Cb^2/\Zb_2$ embedded in the two-instanton moduli space.\footnote{This singular locus can be thought of as parameterizing F-theory configurations where the two D3-branes explore the nonperturbative seven-brane worldvolume as point-like small instantons.} As in \cite{Beem:2019tfp}, we consider an open subset of this locus where $(\mu_{\suf(2)})_{++}\neq0$, which as a Poisson variety can be identified with $T^\ast(\Cb^\ast)$ where $(\mu_{\suf(2)})_{++}$ is the $\Cb^\ast$-valued coordinate and $(\mu_{\suf(2)})_{+-}$ is the cotangent fiber.

From each point in this open subset sprouts the product of two copies of the one-instanton moduli space, which reflects the Higgs branch for the residual IR effective theory.\footnote{The embedding of these one-instanton subspaces is simplified by the fact that the two-instanton moduli space, as a hyperk\"ahler manifold, enjoys an $SU(2)_R\times G\times SU(2)$ isometry group, and at each point on the locus discussed here, this symmetry is broken spontaneously to $SU(2)_{\bar{R}}\times G$ where $SU(2)_{\bar{R}}\cong\text{diag}\left(SU(2)_R\times SU(2)\right)$. Thus the IR R-symmetry can be identified in the UV, which ensures that the one-instanton moduli spaces are genuinely embedded into the two-instanton moduli space rather than only appearing in a scaling region near the singular locus.} We then can construct a dense open subset of the full two-instanton moduli space that has the form of a fibration of those two copies of the one-instanton moduli space over $T^\ast(\Cb^\ast)$. This fibration reflects the indistinguishability of the two one-instanton factors, so we have an open set
\begin{equation}\label{M1cotbundle}
\mathcal{U}=\left(\widetilde{\mathcal M}_{\gf}^{(1)}\times \widetilde{\mathcal M}_{\gf}^{(1)}\times T^{*}(\Cb^*)\right)\big{/} \mathbb{Z}_2~,
\end{equation}
where $ \widetilde{\mathcal M}_{\gf}^{(1)}$ denotes the reduced one instanton moduli space and $\mathbb{Z}_2$ acts by negation on the $\Cb^\ast$ and by exchanging the two one-instanton factors. 

In this patch we can express the generators of the two-instanton coordinate ring in terms of the coordinate ring $\Cb[\mathcal{U}]$. We introduce coordinates (with slightly unconventional names) $(\mathsf{e}^{\frac12},\mathsf{h})$ for $T^{\ast}(\Cb^\ast)$ with their canonical symplectic form $\{\mathsf{h},\mathsf{e}^{\frac12}\}=\mathsf{e}^{\frac12}$, along with two copies $\mathsf{J}_1^A,\,\mathsf{J}_2^A$ of the generators of $\Cb[\widetilde{\MM}^{(1)}_\gf]$ satisfying the Joseph relations \eqref{eqJoseph}. The $\Zb_2$ quotient now acts according to $(\mathsf{h},\mathsf{e}^{\frac{1}{2}},\mathsf{J}_1,\mathsf{J}_2)\mapsto (\mathsf{h},-\mathsf{e}^{\frac{1}{2}},\mathsf{J}_2,\mathsf{J}_1)$. The moment maps of the two-instanton moduli space are now given by
\begin{equation}\label{solvingHiggsBranchrel}
(\mathsf{\mu}_{\suf(2)})_{++}=\mathsf{e}~,\quad(\mathsf{\mu}_{\suf(2)})_{+-}=\tfrac{1}{2}\,\mathsf{h}~,\quad(\mathsf{\mu}_{\suf(2)})_{--}=\big{(}-\mathsf{S}^{\natural}+\tfrac{1}{4}\mathsf{h}^2\big{)}\mathsf{e}^{-1}~,\quad \mu_{\gf}^A=\mathsf{J}_1^{A}+\mathsf{J}_2^{A}~,~~~
\end{equation}
where $\mathsf{S}^{\natural}= \tfrac{1}{4}\kappa_{AB}\mathsf{J}_1^{A}\mathsf{J}_2^{B}$. The additional chiral ring generators can be expressed as
\begin{equation}\label{omegaplusminus}
\omega_{+}^A= \big{(}\mathsf{J}_1^{A}-\mathsf{J}_2^{A}\big{)}\,\mathsf{e}^{\frac{1}{2}}~,\qquad
\omega_{-}^A= \big(-\tfrac{1}{2}if^{A}_{\phantom{A}BC}\,\mathsf{J}_1^{B}\,\mathsf{J}_2^{C}+\tfrac{1}{2}\mathsf{h}\,\left(\mathsf{J}_1^{A}-\mathsf{J}_2^{A}\right) \big)\,\,\mathsf{e}^{-\frac{1}{2}}~.
\end{equation}
In this realization the full complement of Higgs chiral ring relations are solved automatically given that the $\mathsf{J}_i$ satisfy the Joseph relations.

\subsubsection{Affine uplift}
\label{subsubsec:affine_uplift}

Our free field realization will be an ``affine uplift'' of this realization of $\Cb[\MM_{\gf}^{(2)}]$ in terms of $\Cb[\mathcal{U}]$. In particular, we will realize the VOA $\mathcal{V}^{(2)}_{\gf}$ as a vertex operator subalgebra,
\begin{equation}\label{Vgfromfreefields}
\mathcal{V}^{(2)}_{\gf}\,\subset\,\mathcal{V}^{(1)}_{\gf}\otimes\mathcal{V}^{(1)}_{\gf}\otimes \Pi_{\frac{1}{2}}~,
\end{equation}
where $\mathcal{V}^{(1)}_{\gf}=V_{-\frac{1}{6}h^{\vee}-1}(\gf)$ is the associated VOA of the corresponding rank-one SCFT and the VOA $\Pi_{\frac{1}{2}}$ can be expressed in terms of two chiral bosons $\delta(z),\varphi(z)$ with OPEs
\begin{equation}
\label{deltaphiOPE}
\delta(z)\delta(w)\sim \langle \delta,\delta\rangle\,\log (z-w)~,\quad
\varphi(z)\varphi(w)\sim \langle \varphi,\varphi\rangle\,\log (z-w)~,\quad
\delta(z)\varphi(w)\sim0~,
\end{equation}
where $\langle \delta,\delta\rangle=- \langle \varphi,\varphi\rangle$. It is the algebra that includes exponential vertex operators whose lattice momenta are restricted to an isotropic subspace of the full momentum lattice, namely
\begin{equation}\label{PihalfDEF}
\Pi_{\frac{1}{2}}\,:= \,\bigoplus_{\ell=-\infty}^{\infty}\,
\left(V_{\partial\varphi}\otimes V_{\partial\delta}\right)\,e^{\frac{\ell}{2}(\delta+\varphi)}~.
\end{equation}
Equation \eqref{Vgfromfreefields} should be compared to its geometric counterpart \eqref{M1cotbundle}. Once we identify $e^{\delta(z)+\varphi(z)}$ as the VOA avatar of the $\Cb^\ast$-valued $\mathsf{e}$, the generators of $\mathcal{V}^{(2)}_{\gf}$ with non-negative weight under the Cartan of $\mathfrak{su}(2)$ can be immediately written down as
\begin{align}\label{jpp}
j_{++}(z)&=1\otimes e^{\delta(z)+\varphi(z)}~,\\\label{jpm}
j_{+-}(z)&=1\otimes \tfrac{\ksu2}{2}\, \partial \varphi(z)~,\\[3pt]\label{Wplusfreefields}
\mathcal{W}_{+}^A(z)&=\left(\mathcal{J}^A_1-\mathcal{J}^A_2\right)\otimes e^{\frac{1}{2}(\delta(z)+\varphi(z))}~,\\[6pt]
\mathcal{J}^A(z)&=\left(\mathcal{J}^A_1+\mathcal{J}^A_2\right)\otimes 1~.
\end{align}
Here we have fixed the chiral bosons to be normalized according to $\langle \delta,\delta\rangle=-\frac{\ksu2}{2}$, and $\mathcal{J}^A_1$, $\mathcal{J}^A_2$ denote the generators of the two copies of $\mathcal{V}^{(1)}_{\gf}$ in \eqref{Vgfromfreefields}. One can straightforwardly check that the OPEs of these operators correctly reproduce the OPEs given in \eqref{jjOPE}--\eqref{jWOPEs}, \eqref{WWOPEs}. Notice that these expressions in an obvious sense an affinization of \eqref{solvingHiggsBranchrel}, \eqref{omegaplusminus}. To find the remaining generators, it is convenient to first realize, following \cite{Beem:2019tfp}, $j_{--}(z)$ as 
\begin{equation}
\label{jmm}
j_{--}(z)= \Big(-S^{\natural}\otimes 1+1\otimes\left((\tfrac{\ksu2}{2}\,\partial\delta )^2-\tfrac{\ksu2(\ksu2+1)}{2}\partial^2 \delta\right)\Big)
\Big(1\otimes e^{-(\delta+\varphi)}\Big)~.
\end{equation}
The OPEs of the $\mathfrak{su}(2)$ current algebra are then correctly reproduced if and only if the self-OPE of $S^{\natural}$ is as follows: at non-critical level $\ksu2 \neq -2$ the combination $T^{\natural}=S^{\natural}/(\ksu2+2)$ must satisfy the Virasoro OPE with central charge $c^{\natural}=1-6(\ksu2+1)^2/(\ksu2+2)$, while at the critical level, $\ksu2=-2$, $S^{\natural}$, has regular self-OPE. In addition, the requirement that $j_{--}$ commute with the $\fg$ currents $\mathcal{J}^A$ implies that $S^{\natural}$ should do so as well. Within $\mathcal{V}^{(1)}_{\gf}\otimes \mathcal{V}^{(1)}_{\gf}$ there is essentially a unique candidate that can play the role of $S^{\natural}$. It is proportional to the stress tensor for the diagonal coset CFT:
\begin{equation}
\label{Snaturaldef}
S^{\natural}=(\ksu2+2)\,\Big{(}T_1^{\text{Sug}}+T_2^{\text{Sug}}-T_{12}^{\text{Sug}}\Big{)}~,
\end{equation}
where $T_1^{\text{Sug}}$, $T_2^{\text{Sug}}$, $T_{12}^{\text{Sug}}$ are Sugawara stress tensors built using $\mathcal{J}_1^A$, $\mathcal{J}_2^A$, and $\mathcal{J}_{1}^A+\mathcal{J}_{2}^A$, respectively.\footnote{The Sugawara stress tensor of a current algebra generated by $\mathcal{J}$ at level $k$ is given by
\begin{equation}
T^{\text{Sug}}=\frac{1}{2(k+\hv)}\kappa_{AB}\mathcal{J}^A\mathcal{J}^B~,
\end{equation}
where $\hv$ is the dual Coxeter number associated finite-dimensional, simple Lie algebra.}
Notice that in our setup $k_1=k_2=k_{12}/2=k^{\mathfrak{g}}_{2d}/2$. For the $\mathfrak{a}_2$ entry, the definition \eqref{Snaturaldef} looks problematic since $k_{12}+h^{\vee}=-3+3=0$. However, in this case we also have $\ksu2+2=-2+2=0$ resulting in $S^{\natural}$ having contribution only from the (unnormalized) $T_{12}^{\text{Sug}}$ term. 

Having constructed $j_{--}(z)$, we can easily deduce a proposal for the currents $\mathcal{W}_{-}^A(z)$ by considering the OPE of $j_{--}(z)$ with $\mathcal{W}_{+}^A(z)$. We will need to additionally require that a second-order pole is absent, and then the first-order pole will precisely be the desired $\mathcal{W}_{-}^A(z)$. One finds
\begin{equation}
\mathcal{W}_{-}^A(z)=\Big(-\mathcal{U}^A(z)\otimes 1 -(\mathcal{J}_1^A-\mathcal{J}_2^A)\otimes \tfrac{\ksu2}{2}\,\partial\delta(z)\Big)\Big{(}1\otimes e^{-\frac{1}{2}(\delta(z)+\varphi(z))}\Big{)}~,
\end{equation}
where $\mathcal{U}^A$ is defined by the following OPE,
\begin{equation}
S^{\natural}(z)\,(\mathcal{J}_1^A-\mathcal{J}_2^A)(w)\sim \frac{h^{\natural}}{(z-w)^2}(\mathcal{J}_1^A-\mathcal{J}_2^A)(w)+\frac{1}{(z-w)}\mathcal{U}^A(w)~.
\end{equation}
Direct computation then yields 
\begin{equation}
\mathcal{U}^A=\,\tfrac{K}{2}\,i f^A_{BC}\,\mathcal{J}_1^B\,\mathcal{J}_2^C+ \,h^{\natural}\,\partial(\mathcal{J}_1^A-\mathcal{J}_2^A)~,\qquad
h^{\natural}=k^{\fg}\left(\tfrac{\ksu2+2}{k^{\fg}+h^{\vee}}\right)~,
\end{equation}
where $K=-\tfrac{4(2+\ksu2)}{k^{\fg}+h^{\vee}}$. The second-order pole in the $j_{--}\times\mathcal{W}_{+}^A$ OPE vanishes if $\tfrac{1}{2}(\ksu2+\tfrac{1}{2})+h^{\natural}=0$, which can be confirmed to hold for all cases from Table \ref{tab:rank2_deligne}. For these levels, we also find that $K=1$.

At this point all the generators have been constructed and we need to verify that their OPEs close on the algebra that they generate under iterated normally ordered products and derivatives.\footnote{It was noted in Section \ref{sec:explicit_VOAs} that in the $\mathfrak{a}_2$ case there is an additional strong generator identified with the stress tensor. This can be realized as the sum of chiral boson stress tensors and Sugawara stress tensors for the one-instanton factors. Alternatively, this stress tensor appears automatically in the $\mathcal{W}\times\mathcal{W}$ OPE---the VOA is still \emph{generated} by the affine currents and the $\mathcal{W}$ currents, though not strongly.} For this it turns out to be crucial that the levels take the values given in Table \ref{tab:rank2_deligne} and that the currents $\mathcal{J}_1$ and $\mathcal{J}_2$ satisfy the quadratic relations that characterize $\mathcal{V}_\fg^{(1)}$. In particular, the last condition is required for the $\mathfrak{su}(2)$ singlet channel of the $\mathcal{W}\times\mathcal{W}$ OPE to be free of new operators. With all of these conditions satisfied, the OPEs for these free field constructions are identical to the ones give in Section \ref{sec:explicit_VOAs}.

\subsubsection{Higgs branch relations revisited and the \texorpdfstring{$R$}{R}-filtration from free fields.}
\label{subsubsec:Higgs_relations_filtration}

The VOAs we have constructed here exhibit a subtle behavior, previously discussed in \cite{Beem:2017ooy} and observed in examples in \cite{Buican:2017fiq,Beem:2019tfp}, which is that there are Higgs branch relations in the four-dimensional SCFTs that do not have corresponding null vectors in the associated VOA. This phenomenon is closely connected with the nuances of the $R$-filtration on associated VOAs, which has been discussed in detail in \cite{Beem:2017ooy}.

In \cite{Beem:2019tfp} a proposal for the $R$-filtration of a VOA in terms of the corresponding geometric free field realization was put forward. Adopting said proposal, we can study the Higgs branch chiral ring relations collected in \eqref{quadraticrelationHiggs}--\eqref{lastHBrelation} and check that even when they are not realized as null states in the VOA, they are realized at the level of the associated-graded with respect to the $R$-filtration.

We find that in particular, the Higgs branch relations \eqref{quadraticrelationHiggs} and \eqref{NonullrelationHiggs} do not corresponds to null operators in the VOA. The first of these is familiar from many previous investigations, where any time the stress tensor is identified as the Sugawara stress tensor for an affine Kac-Moody (sub-)algebra, there is a ``hidden'' Higgs branch relation. On the other hand, the relation \eqref{NonullrelationHiggs} doesn't appear to be connected to any equally universal phenomenon.

Written in terms of their free field expressions, these composite operators that are supposed to vanish as elements of the Higgs chiral ring take the form\footnote{For future investigations of higher-rank cases, it may be relevant that the quantum numbers of these relations coincide precisely with those of the additional strong generators of the rank-three VOAs that are not Higgs chiral ring generators.}
\begin{eqnarray}\label{TnotnullR}
T_{\mathfrak{su}(2)}^{\text{Sug}}+T_{\fg}^{\text{Sug}}&=&T_{\delta}+T_{\varphi}+T_1^{\text{Sug}}+T_2^{\text{Sug}}~,\\\label{JWnotnullR}
\kappa_{AB}\,\mathcal{J}^A\,\mathcal{W}^B_+&=&\tfrac{5\hv+6}{3}(T_1^{\text{Sug}}+T_2^{\text{Sug}})e^{\frac{1}{2}(\delta+\varphi)}~,
\end{eqnarray}
where in the second equation we have only given the $\mathfrak{su}(2)$ highest weight state for simplicity, and the contribution of the chiral bosons to the stress tensor are defined as
\begin{equation}
T_{\delta}+T_{\varphi}=\tfrac{1}{2}\big(\upsilon_+\upsilon_--\partial \upsilon_-\big)+\tfrac{\ksu2}{2}\,\partial \upsilon_+~,\quad
\upsilon_+\,=\,\partial\left(\delta+\varphi\right)~,\quad
\upsilon_-\,=\,-\tfrac{k}{2}\,\partial\left(\delta-\varphi\right)~.
\end{equation}
While the right hand sides of \eqref{TnotnullR} and \eqref{JWnotnullR} are not null in the VOA, they reside in subspaces of lower-than-expected weight with respect to the $R$-filtration. To be precise, the right-hand sides of \eqref{TnotnullR} and \eqref{JWnotnullR} have weights $R=1$ and $R=3/2$, respectively, which is one less than the sum of weights of the constituents on the left hand sides.\footnote{We recall that the $R$-weights of the chiral bosons are given by $R[\upsilon_+]=0$, $R[\upsilon_-]=1$, $R[e^{\delta+\varphi}]=1$, $R[\partial]=0$, so that $R[T_{\delta}+T_{\varphi}]=1$. The remaining assignments come from the $R$-filtration of the IR VOA, which in this case is $\mathcal{V}^{(1)}_{\fg}\otimes\mathcal{V}^{(1)}_{\fg}$, so that $R[T_1^{\text{Sug}}]=R[T_2^{\text{Sug}}]=1$ \cite{Beem:2019tfp}.} In the associated graded, this leads to the expected Higgs branch relations, while the states on the right-hand side will act as new generators of the resultant commutative algebra. This reinforces the difficulty of constructing the $R$-filtration in an \emph{ad hoc} fashion based on the assignment of $R$-weights to strong generators and correcting on the basis of null states, though in simple cases such a strategy does seem to meet with success \cite{Song:2016yfd}.

\subsection{\texorpdfstring{$H_0$}{H0} theory from Virasoro building blocks}
\label{subsec:H0_FFR}

The free field realization for the rank-two $H_0$ VOA is analogous to the ones presented above, with except that now the building blocks associated to the IR SCFT consist of two copies of the irreducible Virasoro vertex algebra with central charge $c=-22/5$, which will we denote $\text{Vir}_{(2,5)}$.\footnote{The notation originates from the fact that this is the VOA underlying the non-unitary $(2,5)$ minimal model.} This VOA is $C_2$-cofinite, which reflects the fact the Higgs branch of the IR SCFT is a point, so this is a generalized free field realization where we allow $C_2$-cofinite VOAs as elementary building blocks, as proposed in \cite{Beem:2019tfp}. For this example the $\Cb^2/\Zb_2$ subspace where we are studying the low energy effective theory is in fact the generic locus of the Higgs branch. All said, we will therefore be finding an inclusion,
\begin{equation}\label{Vgfromfreefieldsa0}
\mathcal{V}^{(2)}_{\af_0}\,\subset\,\mathcal{V}^{(1)}_{\af_0}\otimes\mathcal{V}^{(1)}_{\af_0}\otimes \Pi_{\frac{1}{2}}~,
\qquad \mathcal{V}^{(1)}_{\af_0}\equiv\text{Vir}_{(2,5)}~.
\end{equation}
We recall that the rank-two $H_0$ VOA is an $\mathfrak{su}(2)$ current algebra at the admissible level $-17/10$ extended by a AKM primary $w_{\alpha}(z)$ of spin $j=\tfrac{1}{2}$ and conformal weight $h=5/2$. The free field realization for the current algebra takes the same form as above (see \eqref{jpp}, \eqref{jpm}, \eqref{jmm}) with 
\begin{equation}
S^{\natural}=(\ksu2+2)\,T^{\natural}\,,\qquad T^{\natural}=T_1+T_2~,
\end{equation}
where $T_1$ and $T_2$ generate the two copies of $\text{Vir}_{(2,5)}$. As in the previous examples, the OPE of the $\suf(2)$ current algebra is correctly reproduced only if $T^{\natural}$ satisfies the Virasoro OPE with central charge $c^{\natural}=1-6(\ksu2+1)^2/(\ksu2+2)$. This is indeed the case with $c^{\natural}=-2\times \frac{22}{5}$ and $\ksu2=-17/10$. A moment's inspection shows that there is a unique candidate AKM primary with the quantum numbers of $w_{+}(z)$ in \eqref{Vgfromfreefieldsa0}. Up to normalization, it is given by
\begin{equation}\label{wplusFORM}
w_{+}(z)=\left(T_1-T_2\right)\otimes e^{\tfrac{1}{2}(\delta+\varphi)}~.
\end{equation}
We can now use the lowering operator $j_{--}(z)$ to construct the lowest-weight state
\begin{equation}
w_{-}(z)=\left(-\tfrac{3}{10}\partial(T_1-T_2)\otimes 1 -(T_1-T_2)\otimes \tfrac{\ksu2}{2} \partial\delta\right)
\left(1 \otimes e^{-\tfrac{1}{2}(\delta+\varphi)}\right)~,
\end{equation}
where the second-order pole cancels precisely fir the relevant values of the level and central charges. With these generators in place one can verify that, up to normalization, the $w_{\alpha}\times w_{\beta}$ OPE takes the form given in \eqref{wwOPEH0}, \eqref{csforH0}.\footnote{The precise normalization \eqref{wplusFORM} corresponds to $c_1=\tfrac{17}{10}\times \tfrac{22}{5}$ in \eqref{wwOPEH0}.}

\section{Rank-two modular equations}
\label{sec:rank_two_LMDE}

As quasi-Lisse VOAs, the vacuum characters of the two-instanton VOAs $\mathcal{V}^{(2)}_{\gf}$ will necessarily be solutions of finite-order linear modular differential equations \cite{Arakawa:2016hkg}. For their rank-one cousins, these differential equations can be expressed in a uniform way as a second-order modular differential operator whose free coefficient is a function of the dual Coxeter number,\footnote{In this section, we use Eisenstein series normalized according to
\begin{equation}
\mathbb{E}_{2k}(\tau) \colonequals -\frac{B_{2k}}{(2k)!}+\frac{2}{(2k-1)!}\sum_{n\geqslant1}\frac{n^{2k-1}q^n}{1-q^n}~,
\end{equation}
where $B_{2k}$ is the $2k$'th Bernoulli number.
}
\begin{equation}\label{eq:rank_one_modular_equation}
\cD^{\text{1-inst}}_{(2)}\goodchi_0^{\text{1-inst},\gf}(q)\colonequals\left(D_q^{(2)} - 5(\hv+1)(\hv-1)\mathbb{E}_4(q)\right)\goodchi_0^{\text{1-inst},\gf}(q)=0~,
\end{equation}
where $D_q^{(n)}=\partial_{(2n-2)}\circ\cdots\circ\partial_{(2)}\circ\partial_{(0)}$ denotes the iterated Serre derivative of modular weight $2n$, with
\begin{equation}
\partial_{(k)}f(q) \colonequals (q\partial_q + k \mathbb{E}_2(\tau))f(q)~.
\end{equation}
One perspective on the uniformity of these differential equations is that these affine current VOAs can be expressed in universal terms, with the nontrivial null states in the vacuum Verma module reflecting the Joseph relations, which as we saw earlier can be described universally within the DC series as the vanishing of the singlet and $\mathbf{Y_2^*}$ representations in the symmetric square of the adjoint. It follows that the null state that leads by recursion to the modular differential equation \eqref{eq:rank_one_modular_equation} should take a universal form.

From this, one might suspect that the rank-two VOAs should, by virtue of their universal form, admit a uniform modular differential equation for their vacuum characters. Indeed, one can determine the leading terms in the vacuum characters directly from the descriptions in Section \ref{sec:explicit_VOAs}, and these admit the uniform expression\footnote{A version of this uniform result has also appeared recently in \cite{Gu:2019dan}.}
\begin{align}\label{eq:2_instanton_uniform}
\goodchi_0^{\text{2-inst},\gf}(q) = q^{\frac{11+5\hv}{24}} \mathrm{PE}\bigg[\frac{1}{1-q}\Big(\big[&(1,\mathbf 1) + (0,\mathbf{Adj})\big]\ q + (\tfrac{1}{2},\mathbf{Adj})\ q^{\frac{3}{2}} - \big[(\tfrac{1}{2},\mathbf{Adj})+(\tfrac{1}{2},\mathbf{Y_2^*})\big]\ q^{\frac{5}{2}} \nn\\
& -\big[(1,\mathbf{1})+(1,\mathbf{Y_2^*}) + (0,\mathbf{Adj})+(0,\mathbf{X_2}) + (0,\mathbf{Y_3^*})\big]\ q^{3} \nn \\
& + (\tfrac{1}{2},\mathbf{Y_2^*})\ q^{\frac{7}{2}} + \ldots\Big)\bigg]~,
\end{align}
where we have utilized the plethystic exponential $\mathrm{PE}[f(x_i)] \colonequals \exp(\sum_{n=1}^{\infty}\frac{1}{n}f(x_i^n))$. In \eqref{eq:2_instanton_uniform} we have indicated the full representation content under $\suf(2)\times\gf$ of each term, but as we are presently interested in the unflavored characters, these expressions should be interpreted as shorthand notation for the dimensions of the indicated finite-dimensional representations. The first several terms of the plethystic exponent manifestly encode the generators and null relations of the vertex operator algebra, while the latter reflect the Higgs branch relations \eqref{quadraticrelationHiggs}--\eqref{lastHBrelation}, except for the relation transforming as $(0,\mathbf 1)$ at order $q^2$, and the relation transforming as $(\frac{1}{2},\mathbf 1)$ at order $q^{\frac{5}{2}}$. Per the previous discussion, these Higgs branch relations do not correspond to null relations of the VOA at the same conformal weight, but rather they can be recovered in the associated graded with respect to the $R$-filtration.

With some additional effort, we can identify the following one-parameter family of fourth-order twisted modular differential equations,\footnote{The term twisted here refers to the fact that the coefficients in the differential operator can be expressed in terms of \emph{twisted} Eisenstein series \cite{Mason:2008zzb}. Alternatively, these are modular with respect to the conjugacy subgroup $\Gamma^{0}(2)\subset PSL(2,\Zb)$.}
\begin{equation}\label{rank2MDE}
\begin{split}
\cD^{\rm 2-inst}_{(4)}=~&D_q^{(4)} + \tfrac{2-\hv}{12}\Theta_{0,1} D_q^{(3)} - \left(\tfrac{25+3\hv+8\hv^2}{288}\Theta_{0,2}+\tfrac{1-9\hv-4\hv^2}{288}\Theta_{1,1}\right)D_q^{(2)}\\
+&\left(\tfrac{138+41\hv-36\hv^2+\hv^3}{6912}\Theta_{0,3}-\tfrac{38+15\hv+20\hv^2-5\hv^3}{2304}\Theta_{1,2}\right)D_q^{(1)}\\
+&\tfrac{(11+5\hv)(11-3\hv-11\hv^2+3\hv^3)}{331776}\Theta_{0,4}+\tfrac{(11+5\hv)(11-51\hv+25\hv^2+9\hv^3)}{82944}\Theta_{1,3}\\
-&\tfrac{167-662\hv+120\hv^2+270\hv^3+65\hv^4}{110592}\Theta_{2,2}~,
\end{split}
\end{equation}
which annihilate the vacuum characters of the rank-two instanton SCFTs.\footnote{One immediately observes a pattern in this differential equation, where the coefficients of an $n$'th order derivative is a polynomial of degree $4-n$ in $\hv$, a pattern which obviously also holds in the rank-one case. Perhaps this pattern will persist at higher rank.} Here $\Theta_{r,s}$ is shorthand notation for the combination of Jacobi theta functions
\begin{equation}
\Theta_{r,s}(\tau)\colonequals \theta_2(\tau)^{4r}\theta_3(\tau)^{4s} + \theta_2(\tau)^{4s}\theta_3(\tau)^{4r}~,\qquad r\leqslant s~,
\end{equation}
and the space of modular forms for the congruence subgroup $\Gamma^0(2)$ is spanned by functions of this type: $M_{2k}(\Gamma^0(2)) = \text{span}\{\Theta_{r,s}(\tau)\ |\ r+s=k \}$. See, for example, the appendix of \cite{Beem:2017ooy} for more details. 

From \eqref{rank2MDE} we can derive the general expression for the ``scaling dimensions'' of the solutions to the $S$-conjugate modular equation, giving
\begin{equation}
\tilde h = \left\{\tfrac{-9-9\hv}{24},\tfrac{-9-5\hv}{24},\tfrac{-9-\hv}{24},\tfrac{15-9\hv}{24}\right\}~,
\end{equation}
which we can use to predict/confirm the $a_{4d}$ Weyl anomaly coefficient using equation 
from which using equations (3.19) from \cite{Beem:2017ooy},
\begin{equation}
a_{4d} =  \frac{\tilde h_{\text{min}}}{2} - \frac{5c_{2d}}{48} =  \frac{23+8\hv}{24}~.
\end{equation}
Indeed, this matches the expression \eqref{a-anomaly} for $n=2$.

\section{Outlook for higher ranks}
\label{sec:four}

We have seen here that the remarkable uniformity of the associated VOAs of the rank-one F-theory SCFTs continues at rank two, which has allowed us to come to grips with the full set of these VOAs quite efficiently. Aside from their intrinsic interest as a diverse family of SCFTs, our analysis here gives one hope that the generalization of this analysis to arbitrary rank may be tractable. To this end, we remark on several observations regarding these higher rank SCFTs and their associated VOAs.

\paragraph{Indices and modular differential equations for $\chi_0^{\mathrm{3-inst}}$.} 

Though we have not established general results for the three-instanton (or higher) theories, we can extract preliminary results for the $\df_4$ theories (which admit Lagrangian descriptions) which should generalize to the rest of the series due to their uniformity. In particular, by explicit computation we have determined the Schur index of the rank-three $\df_4$ theory to high orders in the $q$-expansion 

\small
\begin{equation}
\begin{split}\label{indexD4n3}
\chi_0^{\mathrm{3-inst}\ \df_4}(q) = q^{\frac{80}{24}}(&1 + 31\, q + 60\, q^\frac{3}{2} + 612\, q^2 + 1920\, q^\frac{5}{2} + 10568\, q^3 + 36968\, q^\frac{7}{2} + 157850\, q^4\\
& + 548848\, q^\frac{9}{2}+ 2036655\, q^5 +  6798456\, q^\frac{11}{2} + 22993464\, q^6 + 73082784\, q^\frac{13}{2}\\
& + 230675048\, q^7+ 698674512\, q^\frac{15}{2} + 2086032438\, q^8 + 6042338032\, q^\frac{17}{2}\\
& + 17215132099\, q^9 + 47883383840\, q^\frac{19}{2} + 130994173808 \, q^{10} + \ldots )~.
\end{split}
\end{equation}
\normalsize
One can then verify that this index is annihilated by the following seventh-order modular differential operator,
\small
\begin{equation}
\begin{split}
\cD^{\mathrm{3-inst}\ \df_4}_{(7)}&=~D_q^{(7)} + \frac{41}{120}\Theta_{0,1}D_q^{(6)} + \frac{1}{1440}\left(-7126\ \Theta_{0,2} + 1817\ \Theta_{1,1} \right)D_q^{(5)} \\
&+\frac{1}{5760}\left(-25608\ \Theta_{0,3} + 15881\ \Theta_{1,2} \right)D_q^{(4)} \\
&+\frac{1}{69120}\left(-37904\ \Theta_{0,4} + 999440\  \Theta_{1,3} - 743075\  \Theta_{2,2} \right)D_q^{(3)}\\
&+\frac{1}{165888}\left(3648\ \Theta_{0,5} + 2960696\  \Theta_{1,4} - 2686227\  \Theta_{2,3} \right)D_q^{(2)}\\
&+\frac{1}{5971968}\left(-41344\ \Theta_{0,6} + 22870560\  \Theta_{1,5} +34778766\  \Theta_{2,4} - 58003657\  \Theta_{3,3} \right)D_q^{(1)}\\
&+\frac{1}{71663616}\left(77824\ \Theta_{0,7}-15147776 \ \Theta_{1,6} - 575712384 \ \Theta_{2,5} + 573293501 \ \Theta_{3,4}\right)~.
\end{split}
\end{equation}
\normalsize
This operator implies that the smallest ``scaling dimension'' among the solutions to the $S$-conjugate modular differential equation is given by $\tilde h_{\text{min}}=-\frac{19}{4}$, which in turn predicts an $a_{4d}$ Weyl anomaly coefficient as
\begin{equation}
a_{4d} = \frac{\tilde h_{\text{min}}}{2} - \frac{5c_{2d}}{48}  = \frac{143}{24}~,
\end{equation}
which indeed agrees with \eqref{a-anomaly} for $n=3$ and $\hv=6$.

On this basis, we conjecture that the (unflavored) Schur indices of all rank-three F-theory SCFTs will satisfy a twisted seventh-order modular differential equation whose coefficients are polynomials in the dual Coxeter number. Identifying this universal differential equation might be an interesting starting point for efforts to better understand the rank-three VOAs.

\paragraph{Strong generators for higher rank VOAs $\mathcal{V}^{(n)}_{\fg}$.}

Studying the plethystic logarithm of \eqref{indexD4n3} and its still higher-rank versions leads to a proposal for the set of strong generators of the associated VOAs beyond the Higgs branch generators listed in \eqref{Higgsgen1} and \eqref{Higgsgen2}. In particular, it appears that the higher rank VOAs should be equipped with additional generators with quantum numbers
\begin{equation}\label{NewGENnonHiggs}
\big(\tfrac{\ell}{2},\mathbf{1} \big)_{h = \frac{\ell}{2}+2}~,\qquad\ell=0,1,\dots,n-3~.
\end{equation}
When $n\geqslant 3$ this list includes the stress tensor, which is compatible with the failure of the Sugawara relation between Virasoro central charge and current algebra levels at higher rank. Uniformity within the DC series suggests that this set of strong generators should shared among the higher-rank VOAs for all $\fg$. Indeed, it is tempting to speculate that this is the full list of strong generators of the VOAs $\mathcal{V}^{(n)}_{\fg}$. We hope to put these suggestions to the test in future work.


\acknowledgments

The authors would like to thank Mario Martone and Simone Giacomelli for helpful conversations and useful suggestions. We are pleased to acknowledge the 2019 Pollica summer workshop where some of the final stages of this work were completed, and we are grateful to its supporting organizations: the Simons Foundation (Simons Collaboration on the Nonperturbative Bootstrap) and the INFN. The work of C.B., C.M., and W.P. is partially supported by grant \#494786 from the Simons Foundation. The work of L.R. is partially supported by the NSF grant PHY1620628.

\appendix

\section{Some properties of the Deligne-Cvitanovi\'c exceptional Lie algebras}
\label{app:A}

\subsection{Decomposition of second tensor power of adjoint representation}
\label{subsec:2nd_decomposition}

The Lie algebras of the Deligne series share the property that precisely five real representations appear in the decomposition of the tensor product of two copies of the adjoint representations, three of which occur in the symmetric product and the other two in the antisymmetric product. Following the notations of \cite{cohen1999tensor}, as in the main text, we have
\begin{equation}\label{decompadjxadj}
\mathrm{sym}^2 \mathbf{Adj} = \mathbf{1} + \mathbf{Y_2} + \mathbf{Y_2^*}\;, \qquad \wedge^2 \mathbf{Adj} = \mathbf{Adj} + \mathbf{X_2}~.
\end{equation}
Note that $\suf(2)$ is a degenerate case as the representations $\mathbf{X_2}$ and $\mathbf{Y_2^*}$ are absent (or, more formally, they are identified with the zero-dimensional representation). The dimensions of the various representations entering in \eqref{decompadjxadj} can be expressed uniformly as rational functions of the dual coxeter number $h^\vee$, or, more conveniently, in terms of the parameter $\mu = \frac{6}{h^\vee}$,
\begin{align}
&\dim  \mathbf{Y_2} = -90\frac{(\mu+4)(\mu-5)}{\mu^2(\mu+1)(2\mu+1)}\;, \qquad &&\dim  \mathbf{Y_2^*} = -90\frac{(\mu-3)(\mu+6)}{\mu(\mu+1)^2(2\mu+1)}~,\\
&\dim\mathbf{Adj} = -2 \frac{(\mu-5)(\mu+6)}{\mu(\mu+1)}\;, \qquad &&\dim  \mathbf{X_2} = 5 \frac{(\mu-5)(\mu+6)(\mu-3)(\mu+4)}{\mu^2 (\mu+1)^2}~.
\end{align}
Note that for $\suf(2)$, for which $h^\vee=2$ and thus $\mu = 3$, the dimensions of $\mathbf{X_2}$ and $\mathbf{Y_2^*}$ are indeed zero. For reference, in Table \ref{tab:repsAdjsq}, we give the Dynkin labels of the representations $\mathbf{Adj},\mathbf{Y_2},\mathbf{Y_2^*},$ and $\mathbf{X_2}$ for the full DC series. 
\renewcommand{\arraystretch}{1.4}
\begin{table}[t]
\centering
\begin{tabular}{|c|c|c|c|c|c|c|c|c|}
\hline \hline
 $\mathbf R$  	&  $\af_1$ &  $\af_2$  &  $\gf_2$  &  $\df_4$  &  $\mf{f}_4$  &  $\ef_6$  &  $\ef_7$   &  $\ef_8$\\ 
\hline 
 $\mathbf{Adj}$ &  $[2]$ &  $[11]$  &  $[01]$  &  $[0100]$  &  $[1000]$  &  $[010000]$  &  $[1000000]$   &  $[00000001]$\\ 
\hline
 $\mathbf{Y_2}$ &  $[4]$ &  $[22]$  &  $[02]$  &  $[0200]$  &  $[2000]$  &  $[020000]$  &  $[2000000]$   &  $[00000002]$\\ 
\hline
 $\mathbf{Y_2^*}$ &  --- &  $[11]$  &  $[20]$  & $S_3\cdot[0002]$ &  $[0002]$  &  $[100001]$  &  $[0000010]$   &  $[10000000]$\\ 
\hline
 $\mathbf{X_2}$ &  --- & $\Zb_2\cdot[03]$ &  $[30]$  &  $[1011]$  &  $[0100]$  &  $[000100]$  &  $[0010000]$   &  $[00000010]$\\ 
\hline
\end{tabular}
\caption{\label{tab:repsAdjsq} 
Dynkin labels of representations in $\otimes^2 \mathbf{Adj}$. The outer automorphism $\bZ_2$ of $\af_2$ acts by flipping the two $\af_2$ Dynkin labels, while the $S_3$ outer automorphism of $\df_4$ acts by permuting the first, third, and fourth $\df_4$ Dynkin labels.}
\end{table}

\subsection{Projection operators}
\label{subsec:projection_tensors}

To explicitly construct the rank-two VOAs, we have used behind the scenes the projection operators for each of the representations occurring in $\mathrm{sym}^2 \mathbf{Adj}$. Happily, these have been constructed in the literature \cite{cohen1999tensor}, and we reproduce them here, albeit using slightly different normalization conventions,\footnote{We normalize the Killing form by setting the length squared of the longest root to two. In these conventions, we have in particular that $f^{ABC}f^{EB'C'}\kappa_{BB'}\,\kappa_{CC'} = 2\hv\,\kappa^{AE}$.}
\begin{small}
\begin{align}
&(P_{\mathbf 1})^{BA}_{CD} = \frac{1}{\dim\mathbf{Adj}} \kappa_{CD}\, \kappa^{AB}~,\\
&(P_{\mathbf{Y_2^*}})^{BA}_{CD} = \frac{\mu}{4(1+2\mu)}\left( 2 (\delta^A_D\delta^B_C + \delta^A_C\delta^B_D) + f^{AB}_{\phantom{AB}E} f^{E}_{\phantom{E}CD} +2 f^{AE}_{\phantom{AE}D} f^{B}_{\phantom{B}EC} \right)- \frac{(6+\mu)\kappa_{CD}\, \kappa^{AB}}{(1+2\mu)\dim\mathbf{Adj}} ~, \\
&(P_{\mathbf{Y_2}})^{BA}_{CD} = \frac{1}{2}(\delta^A_D\delta^B_C + \delta^A_C\delta^B_D) - (P_{\mathbf 1})^{BA}_{CD} - (P_{\mathbf{Y_2^*}})^{BA}_{CD}~.
\end{align}
\end{small}
Though we don't need them in our analysis, we also include the projectors onto the representations in the antisymmetric product for completeness,
\begin{align}
&(P_{\mathbf{Adj}})^{BA}_{CD} = -\frac{\mu}{12} f^{AB}_{\phantom{AB}E} f^{E}_{\phantom{E}CD}~,\\
&(P_{\mathbf{X_2}})^{BA}_{CD} = \frac{1}{2}(\delta^A_D\delta^B_C - \delta^A_C\delta^B_D) - (P_{\mathbf{Adj}})^{BA}_{CD}~.
\end{align}
These projectors are idempotent and orthogonal, and their traces equal the dimension of the corresponding representation,
\begin{equation}
(P_{i})^{BA}_{CD} \ (P_{j})^{CD}_{FE} = \delta_{ij}\,(P_{i})^{BA}_{FE}~, \qquad (P_{i})^{AB}_{AB} = \dim \mathbf{R}_i~, \qquad \text{for }i=1,2,\ldots,5~,
\end{equation}
where $\mathbf{R}_i, i=1,\ldots,5$ denote the five representations $\mathbf 1,\mathbf{Y_2},\mathbf{Y_2^*},\mathbf{Adj},\mathbf{X_2}$. 

For the case of $\suf(2)$, the structure constants and Killing form can be chosen to take the explicit form $f^{ABC} = \sqrt{2} \epsilon^{ABC}$ and $\kappa^{AB} = \delta^{AB}$. It is then straightforward to verify that $(P_{\mathbf{Y_2^*}}^{\suf(2)})^{BA}_{CD} \equiv 0$ and $(P_{\mathbf{X_2}}^{\suf(2)})^{BA}_{CD} \equiv 0$, as expected. For $\suf(3)$, the representation $\mathbf{Y_2^*}$ is another copy of the adjoint representation. Its reappearance in the symmetric product is tied to the existence of the cubic Casimir $d^{ABC}$. One may verify that an alternative expression for the projector onto $\mathbf{Y_2^*}$ is as
\begin{equation}
(P_{\mathbf{Y_2^*}}^{\suf(3)})^{BA}_{CD} = \frac{3}{10} d^{ABE} d_{ECD}~.
\end{equation}

\renewcommand{\arraystretch}{1.4}
\begin{table}[t]
\centering
\begin{tabular}{|l|c|c|c|c|c|c|c|c|}
\hline \hline
 ~$\mathbf{R}$~  	&  $\af_1$ &  $\af_2$  &  $\gf_2$  &  $\df_4$  &  $\mf{f}_4$  &  $\ef_6$  &  $\ef_7$   &  $\ef_8$\\ 
\hline 
 ~$\mathbf{A}$ &  --- &  $[22]$  &  $[21]$  &  $S_3\cdot[0102]$  &  $[1002]$  &  $[110001]$  &  $[1000010]$   &  $[10000001]$\\ 
\hline
 ~$\mathbf{Y_3}$ &  $[6]$ &  $[33]$  &  $[03]$  &  $[0300]$  &  $[3000]$  &  $[030000]$  &  $[3000000]$   &  $[00000003]$\\ 
\hline
 ~$\mathbf{Y_3^*}$ &  --- &  $[00]$  &  $[10]$  & $2[0100]$ &  $[0010]$  &  $[100001]$  &  $[0000002]$   &  ---\\ 
\hline
 ~$\mathbf{C}$ &  --- & $\Zb_2\cdot[14]$ &  $[31]$  &  $[1111]$  &  $[1100]$  &  $[010100]$  &  $[1010000]$   &  $[00000011]$\\ 
\hline
 ~$\mathbf{C^*}$ &  $-[2]$ & --- &  $[11]$  &  $2[1011]$  &  $[0011]$  &  $\bZ_2\cdot[000011]$  &  $[0100001]$   &  $[01000000]$\\ 
\hline
 ~$\mathbf{X_3}$ &  $-[4]$ & --- &  $[40]$  &  $S_3\cdot[0022]$  &  $[0020]$  &  $[001010]$  &  $[0001000]$   &  $[00000100]$\\ 
\hline
\end{tabular}
\caption{\label{tab:repsAdj3} 
Dynkin labels of irreducible representations appearing in $\otimes^3 \mathbf{Adj}$. The outer automorphism $\bZ_2$ of $\af_2$ acts by flipping the two Dynkin labels, the $S_3$ outer automorphism of $\df_4$ acts by permuting the first, third and fourth Dynkin labels, and the $\mathbb Z_2$ outer automorphism of $\ef_6$ acts by simultaneously flipping the first and sixth, and the third and fifth Dynkin labels.}
\end{table}

\subsection{Decomposition of third tensor power of adjoint representation}
\label{subsec{third_tensor_power}}

We collect here some data on the representations appearing in the third tensor power of the adjoint representation, see \cite{cohen1996computational} for more details. We will use square brackets to denote plethysms, \eg{}, $[(2)]\mathbf{Adj} = \mathrm{sym}^2 \mathbf{Adj}$ and $[(1,1)]\mathbf{Adj} = \wedge^2 \mathbf{Adj}$. Then one finds
\begin{align}
[(3)]\mathbf{Adj} &= \mathbf{Adj} + \mathbf{X_2} + \mathbf{A} + \mathbf{Y_3} + \mathbf{Y_3^*}~,\\
[(2,1]\mathbf{Adj} &= 2 \mathbf{Adj} + \mathbf{X_2} + \mathbf{Y_2} + \mathbf{Y_2^*} + \mathbf{A} + \mathbf{C} + \mathbf{C^*}~,\\
[(1,1,1)]\mathbf{Adj} &= \mathbf{1} + \mathbf{X_2} + \mathbf{Y_2} + \mathbf{Y_2^*} + \mathbf{X_3}~.
\end{align}
The dimensions of the new representations appearing here can again be written uniformly in terms of $\hv$ or $\mu=\frac{6}{\hv}$ as
\begin{align}
\dim \mathbf{A} &= -27 \frac{(\mu + 4)(\mu + 5)(\mu+6)(\mu-5)(\mu-4)(\mu-3)}{\mu^2(3\mu+1)(3\mu+2)(\mu+1)^2}~,\\
\dim \mathbf{Y_3} &= -10 \frac{(\mu + 4)(\mu + 5)(\mu+6)(5\mu+6)(\mu-5)}{\mu^3(3\mu+1)(2\mu+1)(\mu+1)^2}~,\\
\dim \mathbf{C} &= 640 \frac{(\mu + 3)(\mu + 5)(\mu-5)(\mu-3)}{\mu^3(\mu+1)(2\mu+1)(3\mu+2)}~,\\
\dim \mathbf{X_3} &= -10 \frac{(\mu + 3)(\mu + 5)(\mu+6)(\mu-5)(\mu-4)(\mu-2)}{\mu^3(\mu+1)^3}~.
\end{align}
The dimensions of starred representations are obtained from their unstarred counterparts by the replacement rule $\mu \rightarrow -\mu-1$. We present Dynkin labels for these representations of the various DC algebras in Table \ref{tab:repsAdj3}.

\input{refs.bbl}
\end{document}

%% file: refs.bbl
\providecommand{\href}[2]{#2}\begingroup\raggedright\endgroup